\documentclass[onecolumn, draft, 12pt]{IEEEtran}

\usepackage{algorithmic}
\usepackage{algorithm}
\usepackage[utf8]{inputenc}
\usepackage{graphicx}
\graphicspath{{Figs/}}
\usepackage{amssymb,amsmath,amsthm}
\usepackage{subcaption,siunitx,booktabs}
\usepackage[table]{xcolor}
\usepackage{soul}
\usepackage{diagbox}
\usepackage{cite}
\usepackage{multirow}

\newtheorem{lemma}{Lemma}
\newtheorem{corollary}{Corollary}
\newtheorem{theorem}{Theorem}
\newtheorem{definition}{Definition}
\newtheorem{proposition}{Proposition}

\newtheorem{remark}{Remark}

\newtheorem*{theorem*}{Theorem}
\newtheorem*{lemma*}{Lemma}

\newcommand\blfootnote[1]{%
  \begingroup
  \renewcommand\thefootnote{}\footnote{#1}%
  \addtocounter{footnote}{-1}%
  \endgroup
}

\usepackage{makecell}
\begin{document}

\allowdisplaybreaks

\title{Adaptive relaying for streaming erasure codes in a three node relay network}
\author{Gustavo Kasper Facenda, M. Nikhil Krishnan, Elad Domanovitz, Silas L. Fong, Ashish Khisti, Wai-Tian Tan and John Apostolopoulos \blfootnote{Preliminary results for this paper have been published in \cite{AdaptiveRelay}.}}

\maketitle

\begin{abstract}
    This paper investigates adaptive streaming codes over a three-node relayed network. In this setting, a source node transmits a sequence of message packets to a destination through a relay. The source-to-relay and relay-to-destination links are unreliable and introduce at most $N_1$ and $N_2$ packet erasures, respectively. The destination node must recover each message packet within a strict delay constraint $T$. The paper presents achievable streaming codes for all feasible parameters $\{N_1, N_2, T\}$ that exploit the fact that the relay naturally observes the erasure pattern occurring in the link from source to relay, thus it can adapt its relaying strategy based on these observations. In a recent work, Fong et al. provide streaming codes featuring channel-state-independent relaying strategies. The codes proposed in this paper achieve rates higher than the ones proposed by Fong et al. whenever $N_2 > N_1$, and achieve the same rate when $N_2 = N_1$. The paper also presents an upper bound on the achievable rate that takes into account erasures in both links in order to bound the rate in the second link. The upper bound is shown to be tighter than a trivial bound that considers only the erasures in the second link. 
\end{abstract}
\begin{IEEEkeywords}
	Cloud Computing, Streaming, Low-Latency, Symbol-Wise Decode-and-Forward, Adaptive Relay, Forward Error Correction, Packet Erasure Channel, Relayed Network
\end{IEEEkeywords}

\section{Introduction}\label{intro}
A number of emerging applications including online real-time gaming, real-time video streaming (video conference with multiple users), healthcare (under the name tactile internet), and general augmented reality require efficient low-latency communication. In these applications, data packets are generated at the source in a sequential fashion and must be transmitted to the destination under strict latency constraints. When packets are lost over the network, significant amount of error propagation can occur and suitable methods for error correction are necessary.

There are two main approaches for error correction due to packet losses in communication networks: Automatic repeat request (ARQ) and Forward error correction (FEC). ARQ is {not suitable when considering low latency constraints over long distances, as the round-trip time may be larger than the required delay constraint}. For that reason, FEC schemes are considered more appropriate candidates. The literature has studied codes with strict decoding-delay constraints---called streaming codes---in order to establish fundamental limits of reliable low-latency communication under a variety of packet-loss models. 
Previous works have studied particular, useful cases. In \cite{martinian2004burst}, the authors studied a point-to-point (i.e., two nodes---source and destination) network under a maximal burst erasure pattern. In \cite{leong2012erasure}, the authors have studied, separately, burst erasures and arbitrary erasures. In \cite{badr2013streaming}, the authors have extended the erasure pattern, allowing for both burst erasures and arbitrary erasures. In particular, it was shown that random linear codes \cite{ho2003randomized} are optimal if we are concerned only with correcting arbitrary erasures. Other works that have further studied various aspects of low-latency streaming codes include \cite{JoshiWornell2012, Karzand2017, badr2017layered, badr2017fec, Rashmi2018, krishnan2018rate, fong2019optimal, domanovitz2019explicit, KrishnanLowField2020}.

While most of the prior work on streaming codes has focused on a point-to-point communication link, a network topology that is of practical interest involves a relay node between source and destination, that is, a three-node network. This topology is motivated by numerous applications in which a gateway server, able to decode and encode data, connects two end nodes. Motivated by such considerations, streaming codes for such a setting were first introduced in~\cite{Silas2019}, which derived the time-invariant capacity for the three-node setting, and further extended to a multi-hop network in \cite{domanovitz2020streaming}.

However, the work in \cite{Silas2019} is constrained to time-invariant codes, in particular, the relay does not exploit the knowledge about the erasure pattern it has observed in order to improve its coding scheme. On the other hand, the work in \cite{domanovitz2020streaming} allows for channel adaptation, however, applying the scheme presented in that work in the reduced three-node relayed network does not improve the rate of the streaming code above \cite{Silas2019}, and gains are only observed in the multi-hop setting. 

In our work, we present an achievable coding scheme that is able to outperform the rate achieved by \cite{Silas2019, domanovitz2020streaming} in the three-node relayed network. Furthermore, we present a novel optimization-based upper bound for this setting and an heuristic to find a solution for that optimization.

\subsection{Related Works and Applications}

Our work follows the same adversarial packet erasure channel model used in previous works such as \cite{martinian2004burst, leong2012erasure, badr2013streaming, domanovitz2019explicit, KrishnanLowField2020}. In these works, there is a limit on the number of erasures that may occur, and the goal is to achieve error-free communication within the strict delay constraint. Adaptation of the encoding strategies has been studied in \cite{Cohen2019, Cohen2021} in order to adapt to changing channel statistics. In these works, such adaptation is performed using random linear codes. Relay adaptation has been studied in \cite{domanovitz2020streaming} in order to transmit over a multi-hop setting.

The setting we study, with an intermediate relay between a source and destination, can be used to model communication between a user and a server. In such scenario, it is common that the user communicates with a nearby node that is connected to the same network as the server, and this node then communicates with the server through an internal network. In this case, the link from source to relay models the path from the user to this intermediate node, and the link from relay to destination models the path from it to the server, or vice-versa. In many applications where such a network setting is common, low latency is desirable---frequently, reducing latency is the reason the internal network is built, so the routing can be optimized to reduce delay (e.g. Riot Games' network \cite{maynard-koran_2016} or WTFast network \cite{Hains2020}), rather than the number of hops, which is usually desired by regular Internet Service Providers.

Furthermore, the impact of latency on the user experience in applications such as cloud gaming, where a cloud server performs the computationally intensive tasks such as video rendering, and then transmits only the video output to the player, has been widely studied \cite{Jarschel2013, Quax2013, Clincy2013, Claypool2014, Slivar2014, Wen2014, Schmidt2017}. This latency has different sources, such as propagation delay, hardware delay, server-side processing delay and communications delay, and reducing many of them has been studied, such as server-side delay \cite{lee2014outatime}, video-encoding optimization \cite{Slivar2015} and, as mentioned previously, reducing propagation delay by building internal networks optimized to reduce latency. However, reducing the communication delay seems to be understudied, in particular, the delay caused by packet losses. Considering that the round-trip-delay often represents more than 20\% of the delay budget in these applications \cite{Carrascosa2020}, re-transmissions represent a significant cost in the delay budget. Using streaming codes could make these re-transmissions unnecessary, freeing up a significant fraction of the delay budget.

Similar scenarios appear naturally in other settings, such as virtual and augmented reality, where latency has been linked to motion sickness \cite{Stauffert2020cybersickness}, and, again, e.g. in VR cloud computing, the user communicates with an intermediate node which then communicates with the server.

\section{System Model and Main Results}

\begin{figure}
    \centering
    \includegraphics[draft=false]{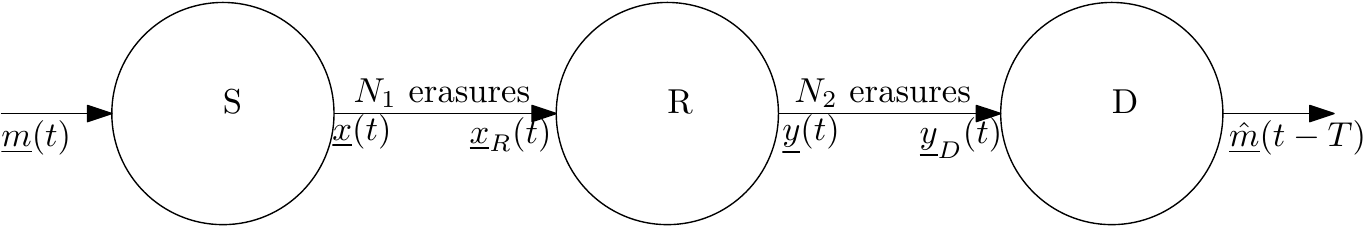}
    \caption{Three node setting}
    \label{fig:networksetting}
\end{figure}

In this section, we formally introduce the problem setting. We use the following notation throughout the paper. The set of non-negative integers is denoted by $\mathbb{Z}_{+}$. The finite field with $q$ elements is denoted by $\mathbb{F}_q$. The set of $l$-dimensional column vectors over $\mathbb{F}_q$ is denoted by $\mathbb{F}_q^l$. For $a,b\in \mathbb{Z}_{+}$, we use $[a:b]$ to denote $\{i\in\mathbb{Z}_{+}\mid a\leq i\leq b\}$. Naturally, we set $[a:\infty]\triangleq \{i\in\mathbb{Z}_{+}\mid i\geq a\}$. 

Consider a three node setup consisting of a source, relay and destination. All packet communication happening in source-to-relay and relay-to-destination links are assumed to be instantaneous, i.e., with no propagation delays. In each discrete time slot $t\in[0:\infty]$, the source has a {\it message packet} $\underline{m}(t)\in\mathbb{F}_q^{k}$ available, which needs to be communicated to the destination via relay. For simplicity, we assume that $\underline{m}(t)\triangleq \underline{0}$, if $t<0$. Towards this, at time-$t$, source invokes a source-side encoder $\mathcal{E}_\text{S}(t):\underbrace{\mathbb{F}_q^{k}\times \cdots \mathbb{F}_q^{k}}_\text{$t+1$ times}\rightarrow \mathbb{F}_q^{n_1}$
to produce a {\it source packet} $\underline{x}(t)\in\mathbb{F}_q^{n_1}$, which is obtained as a function of message packets $\{\underline{m}(t')\}_{t'\in[0:t]}$. Source transmits $\underline{x}(t)$ to the relay over a packet erasure channel. Let $\underline{x}_\text{R}(t)$ denote the packet received by relay. We have:  
\begin{align}
    \underline{x}_\text{R}(t)  =  \left\{ \begin{array}{cl} *, & \text{if $\underline{x}(t)$ is erased}, \\
	\underline{x}(t), & \text{otherwise}. \end{array} \right.
\end{align}

In time-$t$, once relay receives $\underline{x}_\text{R}(t)$, it produces a {\it relay packet} $\underline{y}(t)\in\mathbb{F}_q^{n_2}$ by invoking a relay-side encoder: 
\begin{equation}
\mathcal{E}_\text{R}(t):
\underbrace{\mathbb{F}_q^{n_1}\cup\{*\}\times \cdots \times \mathbb{F}_q^{n_1}\cup\{*\}}_\text{$t+1$ times}\rightarrow \mathbb{F}_q^{n_2}.
\end{equation}
 The relay packet $\underline{y}(t)$ is a function of packets $\{\underline{x}_\text{R}(t')\}_{t'\in[0:t]}$. Relay transmits $\underline{y}(t)$ to the destination in time-$t$. Owing to erasures in relay-to-destination link, the packet $\underline{y}_\text{D}(t)$ received by destination in time-$t$ is given by:
\begin{align}
\underline{y}_\text{D}(t) = \left\{ \begin{array}{cl} *, & \text{if $\underline{y}(t)$ is erased}, \\
	\underline{y}(t), & \text{otherwise}. \end{array} \right.
\end{align}

At time-$(t+T)$, destination uses decoder:
\begin{equation}
\mathcal{D}(t):\underbrace{\mathbb{F}_q^{n_2}\cup\{*\}\times \cdots \times \mathbb{F}_q^{n_2}\cup\{*\}}_\text{$t+1+T$ times}\rightarrow \mathbb{F}_q^{k} 
\end{equation}
to obtain an estimate $\underline{\hat{m}}(t)\in\mathbb{F}_q^{k\times 1}$ of $\underline{m}(t)$ as a function of received packets $\{\underline{y}_\text{D}(t')\}_{t'\in[0:t+T]}$. The decoder is delay-constrained as $\underline{m}(t)$ has to be estimated by time-$(t+T)$. The tuple $(\{\mathcal{E}_\text{S}(t)\},\{\mathcal{E}_\text{R}(t)\},\{\mathcal{D}(t)\})$ constitutes an {\it $(n_1,n_2,k,T)_q$ streaming code}. This setting is illustrated in Fig.~\ref{fig:networksetting}.

\begin{definition}[Rate of a $(n_1, n_2, k, T)_q$ streaming code]
    The rate of an $(n_1,n_2,k,T)_q$ streaming code is defined to be $\frac{k}{\max\{n_1,n_2\}}$.
\end{definition}

\begin{definition}[Erasure Sequences]\normalfont
	A source-relay erasure sequence denoted by $e_\text{S}^\infty\triangleq \{e_{\text{S},t}\}_{t\in[0:\infty]}$ is a binary sequence, where $e_{\text{S},t}=1$ iff ${x}_\text{R}(t)=*$. Similarly, a relay-destination erasure sequence  $e_\text{R}^\infty\triangleq \{e_{\text{R},t}\}_{t\in[0:\infty]}$ will have  $e_{\text{R},t}=1$ iff ${y}_\text{D}(t)=*$
\end{definition}

\begin{definition}[$N$-Erasure Sequences]\normalfont
	Let $N\in \mathbb{Z}_+$. A source-relay erasure sequence $e_\text{S}^\infty$ is defined to be an $N$-erasure sequence if $\sum_{t\in[0:\infty]}e_{\text{S},t}\leq N$. Similarly, $e_\text{R}^\infty$ is an $N$-erasure sequence if $\sum_{t\in[0:\infty]}e_{\text{R},t}\leq N$.
\end{definition}

\begin{definition}[$(N_1,N_2, T)$-Achievability]\normalfont
	An $(n_1,n_2,k,T)_q$ streaming code is defined to be $(N_1,N_2, T)$-achievable if it is possible to perfectly reconstruct all message packets (i.e., $\underline{\hat{m}}(t)=\underline{m}(t)$ for all $t$) at the destination in presence of (i) any $N_1$-erasure sequence $e_\text{S}^\infty$ and (ii) any $N_2$-erasure sequence $e_\text{R}^\infty$. Similarly, a rate $R$ is said to be $(N_1, N_2, T)$-achievable if there exists an $(n_1, n_2, k, T)_q$ code such that: the code is $(N_1, N_2, T)$-achievable and $R = \frac{k}{\max(n_1, n_2)}$. \label{definition:achievability}
\end{definition}
%This is because, if $N_1+N_2>T$, in presence of erasure of erasure patterns $e_\text{S}^\infty=\{\underbrace{0,\ldots,0}_{i},\underbrace{1\ldots,1}_{N_1},0,\ldots\}$ and $e_\text{R}^\infty=\{\underbrace{0,\ldots,0}_{i+N_1},\underbrace{1\ldots,1}_{N_2},0,\ldots\}$, it is impossible for the destination to recover $\underline{m}(i)$ by time-$(i+T)$. 
\begin{definition}[$(N_1, N_2, T)$-Capacity]
    The $(N_1, N_2, T)$-capacity, denoted as $C_{N_1, N_2, T}$, is the maximum of all rates that are $(N_1, N_2, T)$-achievable, as defined in Definition~\ref{definition:achievability}. 
\end{definition}
It may be noted that, if $N_1 + N_2 > T$, the $(N_1, N_2, T)$-capacity is 0.

\begin{remark} \label{remark:recover}
	Error protection provided by $(N_1,N_2, T)$-achievable $(n_1,n_2,k,T)$ streaming codes may appear to be limiting, as they consider only $N_1$ erasures across all time slots $[0:\infty]$ in source-relay link and $N_2$ erasures across all time slots $[0:\infty]$ in relay-destination link. However, owing to the delay-constrained decoder, these codes can in fact recover from any $e_S^\infty$, $e_R^\infty$ which satisfy: $\sum_{t'=i}^{i+T}e_{S,t}\leq N_1$ and $\sum_{t'=i}^{i+T}e_{R,t}\leq N_2$  for all $i\in[0:\infty]$. i.e., in any sliding window of $T+1$ consecutive time slots, source-relay and relay-destination links see at most $N_1$ and $N_2$ erasures, respectively.
\end{remark}

\section{Proposed Scheme}

In this section, we present the code construction and the achievable rate of such streaming code. We start by stating our main results, then, we present the main ideas that lead to these results, a toy-case example demonstrating the key ideas, and finally the general code construction. 

\begin{theorem}\label{theorem:achievable}
    For any $N_1$, $N_2$ and $T$, there exists an $(N_1, N_2, T)$-achievable $(n_1, n_2, k, T)_q$ streaming code with rate \mbox{$R = \min \left( R_1, R_2  \right)$} where 
    \begin{align}
        %R &= \min \left( R_1, R_2  \right) \\
        R_1 &= \frac{T + 1 - N_1 - N_2}{T + 1 - N_2} \\
        R_2 &= \frac{T + 1 - N_2}{T + 1 + \sum_{i = 0}^{N_1} \frac{N_1 - i}{T + 1 - N_2 - (N_1 - i)} + \delta}
    \end{align}
    and where $\delta$ is an overhead bounded by $\frac{1}{c} \lceil {(T+1)} \log_q 2 \rceil$, where $c$ is an arbitrary integer constant, thus the overhead goes to 0 as $c$ increases.
\end{theorem}
\begin{remark}
    In Theorem~\ref{theorem:achievable}, the $\delta$ term represents the fact that, since the relay is changing its coding strategy according to the erasure pattern that has occurred in the link from source to relay, that erasure pattern must also be informed to the destination by the relay. 
    %However, as that information depends only on the erasure pattern, it can be seen that, by increasing the packet length, this overhead becomes negligible. For that reason, 
    In order to keep it simpler, we present our examples and highlight the main ideas assuming the destination is given the erasure pattern that has occurred from source to relay. We also present a naive solution to transmit this necessary information, which is independent of the packet size, thus, by making the code infinitely large, this overhead is negligible. In the theorem, the constant $c$ represents the number of codewords we multiplex in the same packet, thus increasing it increases the packet size and makes the overhead negligible. More details are given in the end of Section~\ref{sec:relay_destination_coding}. 
\end{remark}

\begin{corollary}
    For any $T$ and $N_2 > N_1$, for a sufficiently large $q$, there exists an $(N_1, N_2, T)$-achievable channel-state-dependent $(n_1, n_2, k, T)_q$ streaming code that achieves a rate (strictly) higher than $R = \frac{T + 1 - N_1 - N_2}{T + 1 - N_1}$ which is the rate achieved by channel-state-independent $(N_1, N_2, T)$-achievable streaming codes \cite{Silas2019}. 
\end{corollary}

\subsection{Main Ideas}

As mentioned in previous sections, our streaming code differs from the one presented in \cite{Silas2019} mainly in its relaying function. The encoding function from source to relay follows the same core idea. However, the relaying function differs greatly by employing the knowledge of the erasure pattern that has occurred. Below, we highlight the main concepts which allow us to exploit that knowledge.

\subsubsection{Packet-wise variable rate}

In our coding scheme, from relay to destination, each message packet is transmitted using a different rate, based on the erasure pattern its respective source packet has been subject to in the link from source to relay. In \cite{Silas2019}, each message is transmitted assuming it has been subject to the worst case erasure pattern, which leads to every message being transmitted with an effective delay of $T' = T - N_1$ from relay to destination. In other words, in prior works, information about a message packet $\underline{m}(t)$ is only relayed starting at time $t + N_1$, even though it might have been recovered earlier than that, which occurs, for example, if the respective source packet has not been erased. This implies that the respective message packet can be transmitted with effective delay equal to $T' = T$, that is, the source-to-relay link has introduced no delay to that message packet. Furthermore, from our channel model assumptions, some source packets are guaranteed to not be erased (otherwise the capacity is 0), thus, some message packets can certainly be transmitted from relay to destination with effective delay $T' = T$. Since these packets can be transmitted with a higher effective delay, they can be transmitted with a higher rate, which will lead to a higher overall rate, as we will see later. This also leads to a variable streaming code rate, however, we handle that by zero padding.

\subsubsection{Within-message variable rate}

Not only each message packet is transmitted with a different rate according to the erasure pattern it is subject to in the first link, this rate is also adapted on the fly as new erasures occur. For example, consider source packet $\underline{x}(t)$ has been erased. At time $t$, message packet $\underline{m}(t)$ is subject to one erasure. Then, assume we successfully recover $\underline{x}(t + 1)$. The relay start transmitting some information about $\underline{m}(t)$. Up to this point, only one erasure has occurred, and we transmit accordingly. Then, assume $\underline{x}(t+2)$ is again erased. Now, $\underline{m}(t)$ is subject to \textit{two} erasures, rather than one, and its rate of transmission should adapt accordingly to a lower rate. We achieve this rate adaptation by using long maximum-distance separable (MDS) codes, which encode across all $k$ message symbols and allow for variable-rate transmission, rather than diagonally-interleaved ``short'' MDS codes, which were used in prior works.

\subsubsection{Noisy Relaying}

Another key difference is that, in \cite{Silas2019}, the concept of ``symbol-wise decode-and-forward'' was introduced, where the relay decodes symbols of the message packet and then forwards them. However, the relay is only able to forward information about symbols it has been able to fully recover. In our coding scheme, the relay transmits \textit{noisy} symbols, i.e., symbols that contain interference caused by \textit{past} message packets. Because the destination is required to recover packets sequentially, interference of past messages can be cancelled by the deadline at the destination.

\subsection{Example}

Consider, for example, a network with $N_1 = 2$, $N_2 = 3$ and $T = 6$. Let us consider $k = 24$, that is, each message packet consists of 24 symbols. We denote by $m_i(t)$ the $i$th message symbol at time $t$. We use the notation $a : i : b = \{a, a + i, a + 2i, \ldots, b\}$ and $m_{a : i : b}(t) = \begin{bmatrix} m_a(t), m_{a + i}(t), \ldots, m_{b}(t) \end{bmatrix}$.

As we show next, the source is using the same coding scheme as \cite{Silas2019} which amounts to a systematic transmission with diagonally interleaved block codes which results in transmitting at rate of $24/48=1/2$.

First, let us consider the scenario where the erasures in the first link occur in a burst. As can be seen in Table~\ref{tab:burster}, the relay starts transmitting source packets that have not been erased immediately with rate $4/7$. This can be seen, for example, for $m(3)$. On the other hand, source packets that are subject to erasures are transmitted with lower rate, since they need to be transmitted ``faster'' (i.e., with a smaller effective delay). Thus, it can be seen that $\underline{m}(4)$ is transmitted with rate $3/6$ and $\underline{m}(5)$ is transmitted with rate $2/5$. This highlights the packet-wise variable rate aspect of our coding strategy. Further, note that the relay is unable to recover the symbols $m_{2:2:16}(5)$ at time 6. If the relay would wait until it can be recovered without the interference from $\underline{m}(4)$, then $\underline{m}(5)$ would also need to be transmitted with rate $2/5$. In order to be able to transmit it with a better rate, we instead transmit the noisy symbols $m_{2:2:16}'(5) = m_{2:2:16}(5) + m_{1:2:15}(4)$. Then, at the destination, since it can recover $\underline{m}(4)$ entirely at time 10, it can cancel out the interference at time 11 and recover $\underline{m}(5)$. This highlights the concept of noisy relaying.

% \begin{figure}[h]
%     \centering
%     \hspace*{-0.15in}
%     \includegraphics[scale=0.45, draft=false]{example1}
%     \caption{$T=6,~N_1=2~N_2=3$ example of burst erasures in the link between source and relay} \label{fig:burster}
% \end{figure}

\begin{table}
\centering
\caption{Example of the proposed encoding scheme in case of burst erasures in the link between source and relay for $T = 6$, $N_1 = 2$, $N_2 = 3$.}\label{tab:burster}
\resizebox{\columnwidth}{!}{
\begin{tabular}{|c|c|c|c|c|c|c|c|c|c|}
\hline
Time                    & 3                                & \cellcolor[HTML]{FFCCC9} 4                                & \cellcolor[HTML]{FFCCC9}5                                & 6                                & 7                                & 8                                & 9                                & 10                               & 11                               \\ \hline
 & $m_{1:2:24}(3)$                  & {\color[HTML]{6434FC}  \cellcolor[HTML]{FFCCC9}$m_{1:2:24}(4)$}                  & \cellcolor[HTML]{FFCCC9}{\color[HTML]{32CB00}$m_{1:2:24}(5)$}                  & $m_{1:2:24}(6)$                  & $m_{1:2:24}(7)$                  & $m_{1:2:24}(8)$                  & $m_{1:2:24}(9)$                  & $m_{1:2:24}(10)$                 & $m_{1:2:24}(11)$                 \\ \cline{2-10} 
  \multirow{4}{*}{\rotatebox[origin=c]{90}{Source}}                      & $m_{2:2:24}(3)$                  & \cellcolor[HTML]{FFCCC9} $m_{2:2:24}(4)$                  & {\color[HTML]{6434FC} \cellcolor[HTML]{FFCCC9} $m_{2:2:24}(5)$}                  & {\color[HTML]{32CB00}$m_{2:2:24}(6)$}                  & $m_{2:2:24}(7)$                  & $m_{2:2:24}(8)$                  & $m_{2:2:24}(9)$                  & $m_{2:2:24}(10)$                 & $m_{2:2:24}(11)$                 \\ \cline{2-10} 
                        &  \makecell{$m_{1:2:24}(1)$\\ $+ m_{2:2:24}(2)$}  & \cellcolor[HTML]{FFCCC9} \makecell{$m_{1:2:24}(2)$ \\$ + m_{2:2:24}(3)$  }& \cellcolor[HTML]{FFCCC9} \makecell{$m_{1:2:24}(3)$\\$ + m_{2:2:24}(4)$}  & \makecell{{\color[HTML]{6434FC} $m_{1:2:24}(4)$}\\{\color[HTML]{6434FC}$  + m_{2:2:24}(5)$}}  & {\color[HTML]{32CB00}\makecell{$m_{1:2:24}(5)$\\$ + m_{2:2:24}(6)$}}  & \makecell{$m_{1:2:24}(6)$\\$ + m_{2:2:24}(7)$}  & \makecell{$m_{1:2:24}(7)$\\$ + m_{2:2:24}(8)$}  & \makecell{$m_{1:2:24}(8)$\\$ + m_{2:2:24}(9)$}  & \makecell{$m_{1:2:24}(9)$\\$ + m_{2:2:24}(10)$} \\ \cline{2-10} 
                        & \makecell{$m_{1:2:24}(0)$\\$ + 2m_{2:2:24}(1)$} & \cellcolor[HTML]{FFCCC9} \makecell{$m_{1:2:24}(1)$\\$ + 2m_{2:2:24}(2)$} & \cellcolor[HTML]{FFCCC9} \makecell{$m_{1:2:24}(2)$\\$ + 2m_{2:2:24}(3)$} & \makecell{$m_{1:2:24}(3)$\\$ + 2m_{2:2:24}(4)$} & \makecell{{\color[HTML]{6434FC}$m_{1:2:24}(4)$}\\{\color[HTML]{6434FC}$ + 2m_{2:2:24}(5)$}} & {\color[HTML]{32CB00}\makecell{$m_{1:2:24}(5)$\\$ + 2m_{2:2:24}(6)$}} & \makecell{$m_{1:2:24}(6)$\\$ + 2m_{2:2:24}(7)$} & \makecell{$m_{1:2:24}(7)$\\$ + 2m_{2:2:24}(8)$} & \makecell{$m_{1:2:24}(8) $\\$+ 2m_{2:2:24}(9)$} \\ \hline
\end{tabular}}

\vspace{0.2cm}

\resizebox{\columnwidth}{!}{
\begin{tabular}{|c|c|c|c|c|c|c|c|c|c|}
\hline
Time                    & 3                                      & 4                                      & 5                                      & {\color[HTML]{333333} 6}                                                                                        & 7                                           & 8                                                                                                  & 9                                                                                                   & 10                                                                                                   & 11                                   \\ \hline
                        & {\color[HTML]{6434FC} $m_{1:4:24}(3)$} &                                        &                                        & $m_{1:4:24}(6)$                                                                                                 & $m_{1:4:24}(7)$                             & $m_{1:4:24}(8)$                                                                                    & $m_{1:4:24}(9)$                                                                                     & $m_{1:4:24}(10)$                                                                                     & $m_{1:4:24}(11)$                     \\ \cline{2-10} 
                        & $m_{2:4:24}(2)$                        & {\color[HTML]{6434FC} $m_{2:4:24}(3)$} &                                        &                                                                                                                 & $m_{2:4:24}(6)$                             & $m_{2:4:24}(7)$                                                                                    & $m_{2:4:24}(8)$                                                                                     & $m_{2:4:24}(9)$                                                                                      & $m_{2:4:24}(10)$                     \\ \cline{2-10} 
                        & $m_{3:4:24}(1)$                        & $m_{3:4:24}(2)$                        & {\color[HTML]{6434FC} $m_{3:4:24}(3)$} &                                                                                                                 &                                             & $m_{3:4:24}(6)$                                                                                    & $m_{3:4:24}(7)$                                                                                     & $m_{3:4:24}(8)$                                                                                      & $m_{3:4:24}(9)$                      \\ \cline{2-10} 
                        & $m_{4:4:24}(0)$                        & $m_{4:4:24}(1)$                        & $m_{4:4:24}(2)$                        & {\color[HTML]{6434FC} $m_{4:4:24}(3)$}                                                                          &                                             &                                                                                                    & $m_{4:4:24}(6)$                                                                                     & $m_{4:4:24}(7)$                                                                                      & $m_{4:4:24}(8)$                      \\ \cline{2-10} 
                        &                                        & $p^{(0)}_{1:1:6}(4)$                   & $p^{(1)}_{1:1:6}(5)$                   & $p^{(2)}_{1:1:6}(6)$                                                                                            & {\color[HTML]{6434FC} $p^{(3)}_{1:1:6}(7)$} &                                                                                                    &                                                                                                     & $p^{(6)}_{1:1:6}(10)$                                                                                & $p^{(7)}_{1:1:6}(11)$                \\ \cline{2-10} 
                        &                                        &                                        & $p^{(0)}_{1:1:6}(5)$                   & $p^{(1)}_{1:1:6}(6)$                                                                                            & $p^{(2)}_{1:1:6}(7)$                        & {\color[HTML]{6434FC} $p^{(3)}_{1:1:6}(8)$}                                                        &                                                                                                     &                                                                                                      & $p^{(6)}_{1:1:6}(11)$                \\ \cline{2-10} 
                        &                                        &                                        &                                        & $p^{(0)}_{1:1:6}(6)$                                                                                            & $p^{(1)}_{1:1:6}(7)$                        & $p^{(2)}_{1:1:6}(8)$                                                                               & {\color[HTML]{6434FC} $p^{(3)}_{1:1:6}(9)$}                                                         &                                                                                                      &                                      \\  \cline{2-10} \multirow{-9}{*}{\rotatebox[origin=c]{90}{Relay}}
                        &                                        &                                        &                                        & {\color[HTML]{FE0000} $m_{2:2:24}$}                                                                             & {\color[HTML]{FE0000} $m_{1:2:24}$}         & {\color[HTML]{FE0000} \makecell{$p^{(4)}_{1:1:12}(8)$ \\ $ = m_{2:2:24}(4)$ \\ $+ m_{1:2:24}(4)$}} & {\color[HTML]{FE0000} \makecell{$p^{(4)}_{1:1:12}(9)$ \\ $ = m_{2:2:24}(4)$ \\ $+ 2m_{1:2:24}(4)$}} & {\color[HTML]{FE0000} \makecell{$p^{(4)}_{1:1:12}(10)$ \\ $ = m_{2:2:24}(4)$ \\ $+ 3m_{1:2:24}(4)$}} &                                      \\ \cline{2-10}  
 &                                        &                                        &                                        & \cellcolor[HTML]{FFFFC7}{\color[HTML]{32CB00} \makecell{$m'_{2:2:16}(5) = m_{1:2:16}(4)$ \\ $+ m_{2:2:16}(5)$}} & {\color[HTML]{32CB00} $m_{1:2:16}(5)$}      & {\color[HTML]{32CB00} $m_{17:1:24}(5)$}                                                            & {\color[HTML]{32CB00} $p^{(5)}(9)$}                                                                 & {\color[HTML]{32CB00} $p^{(5)}(10)$}                                                                 & {\color[HTML]{32CB00} $p^{(5)}(11)$} \\ \hline
\end{tabular}}

\end{table}

Let us now consider the scenario where the erasures are spaced, as shown in Table~\ref{tab:spaceder}. Since $\underline{m}(4)$ is subject to only one erasure, we attempt to transmit it with rate $3/6$, similar to how the packet $\underline{m}(5)$ was transmitted in the previous example. However, another erasure occurs at time 6 and therefore the relay does not have enough symbols to keep transmitting with such high rate. Instead, we simply transmit the remaining symbols that have been previously recovered at time 5. Then, at time 7, we start transmitting $\underline{m}(6)$ with rate $3/6$, and we change the rate used for $\underline{m}(4)$ from $3/6$ to $2/5$, as it now has been subject to two erasures. This highlights the concept of within-message variable rate.

% \begin{figure}[h]
%     \centering
%     \hspace*{-0.15in}
%     \includegraphics[scale=0.45, draft=false]{example2}
%     \caption{$T=6,~N_1=2~N_2=3$ example of spaced erasures in the link between source and relay} \label{fig:spaceder}
% \end{figure}

\begin{table}
\centering
\caption{Example of the proposed encoding scheme in case of spaced erasures in the link between source and relay for $T = 6$, $N_1 = 2$, $N_2 = 3$.} \label{tab:spaceder}
\resizebox{\columnwidth}{!}{
\begin{tabular}{|c|c|c|c|c|c|c|c|c|c|}
\hline
Time                    & 3                                & \cellcolor[HTML]{FFCCC9} 4                                & 5                                & \cellcolor[HTML]{FFCCC9}6                                & 7                                & 8                                & 9                                & 10                               & 11                               \\ \hline
 & $m_{1:2:24}(3)$                  & {\color[HTML]{6434FC}  \cellcolor[HTML]{FFCCC9}$m_{1:2:24}(4)$}                  & {\color[HTML]{32CB00}$m_{1:2:24}(5)$}                  & \cellcolor[HTML]{FFCCC9} $m_{1:2:24}(6)$                  & $m_{1:2:24}(7)$                  & $m_{1:2:24}(8)$                  & $m_{1:2:24}(9)$                  & $m_{1:2:24}(10)$                 & $m_{1:2:24}(11)$                 \\ \cline{2-10} 
  \multirow{4}{*}{\rotatebox[origin=c]{90}{Source}}                      & $m_{2:2:24}(3)$                  & \cellcolor[HTML]{FFCCC9} $m_{2:2:24}(4)$                  & {\color[HTML]{6434FC}  $m_{2:2:24}(5)$}                  & \cellcolor[HTML]{FFCCC9}{\color[HTML]{32CB00}$m_{2:2:24}(6)$}                  & $m_{2:2:24}(7)$                  & $m_{2:2:24}(8)$                  & $m_{2:2:24}(9)$                  & $m_{2:2:24}(10)$                 & $m_{2:2:24}(11)$                 \\ \cline{2-10} 
                        &  \makecell{$m_{1:2:24}(1)$\\ $+ m_{2:2:24}(2)$}  & \cellcolor[HTML]{FFCCC9} \makecell{$m_{1:2:24}(2)$ \\$ + m_{2:2:24}(3)$  }&  \makecell{$m_{1:2:24}(3)$\\$ + m_{2:2:24}(4)$}  &\cellcolor[HTML]{FFCCC9} \makecell{{\color[HTML]{6434FC} $m_{1:2:24}(4)$}\\{\color[HTML]{6434FC}$  + m_{2:2:24}(5)$}}  & {\color[HTML]{32CB00}\makecell{$m_{1:2:24}(5)$\\$ + m_{2:2:24}(6)$}}  & \makecell{$m_{1:2:24}(6)$\\$ + m_{2:2:24}(7)$}  & \makecell{$m_{1:2:24}(7)$\\$ + m_{2:2:24}(8)$}  & \makecell{$m_{1:2:24}(8)$\\$ + m_{2:2:24}(9)$}  & \makecell{$m_{1:2:24}(9)$\\$ + m_{2:2:24}(10)$} \\ \cline{2-10} 
                        & \makecell{$m_{1:2:24}(0)$\\$ + 2m_{2:2:24}(1)$} & \cellcolor[HTML]{FFCCC9} \makecell{$m_{1:2:24}(1)$\\$ + 2m_{2:2:24}(2)$} &  \makecell{$m_{1:2:24}(2)$\\$ + 2m_{2:2:24}(3)$} & \cellcolor[HTML]{FFCCC9}\makecell{$m_{1:2:24}(3)$\\$ + 2m_{2:2:24}(4)$} & \makecell{{\color[HTML]{6434FC}$m_{1:2:24}(4)$}\\{\color[HTML]{6434FC}$ + 2m_{2:2:24}(5)$}} & {\color[HTML]{32CB00}\makecell{$m_{1:2:24}(5)$\\$ + 2m_{2:2:24}(6)$}} & \makecell{$m_{1:2:24}(6)$\\$ + 2m_{2:2:24}(7)$} & \makecell{$m_{1:2:24}(7)$\\$ + 2m_{2:2:24}(8)$} & \makecell{$m_{1:2:24}(8) $\\$+ 2m_{2:2:24}(9)$} \\ \hline
\end{tabular}}

\vspace{0.2cm}

\resizebox{\columnwidth}{!}{
\begin{tabular}{|c|c|c|c|c|c|c|c|c|c|c|}
\hline
Time                     & 3                                      & 4                                      & 5                                      & {\color[HTML]{333333} 6}                & 7                                           & 8                                                                                                 & 9                                                                                                  & 10                                                                                                & 11                                   & 12                                   \\ \hline
                         & {\color[HTML]{6434FC} $m_{1:4:24}(3)$} &                                        & $m_{1:4:24}(5)$                        &                                         & $m_{1:4:24}(7)$                             & $m_{1:4:24}(8)$                                                                                   & $m_{1:4:24}(9)$                                                                                    & $m_{1:4:24}(10)$                                                                                  & $m_{1:4:24}(11)$                     &                                      \\ \cline{2-11} 
                         & $m_{2:4:24}(2)$                        & {\color[HTML]{6434FC} $m_{2:4:24}(3)$} &                                        & $m_{2:4:24}(5)$                         &                                             & $m_{2:4:24}(7)$                                                                                   & $m_{2:4:24}(8)$                                                                                    & $m_{2:4:24}(9)$                                                                                   & $m_{2:4:24}(10)$                     &                                      \\ \cline{2-11} 
                         & $m_{3:4:24}(1)$                        & $m_{3:4:24}(2)$                        & {\color[HTML]{6434FC} $m_{3:4:24}(3)$} &                                         & $m_{3:4:24}(5)$                             &                                                                                                   & $m_{3:4:24}(7)$                                                                                    & $m_{3:4:24}(8)$                                                                                   & $m_{3:4:24}(9)$                      &                                      \\ \cline{2-11} 
                         & $m_{4:4:24}(0)$                        & $m_{4:4:24}(1)$                        & $m_{4:4:24}(2)$                        & {\color[HTML]{6434FC} $m_{4:4:24}(3)$}  &                                             & $m_{4:4:24}(5)$                                                                                   &                                                                                                    & $m_{4:4:24}(7)$                                                                                   & $m_{4:4:24}(8)$                      &                                      \\ \cline{2-11} 
                         &                                        & $p^{(0)}_{1:1:6}(4)$                   & $p^{(1)}_{1:1:6}(5)$                   & $p^{(2)}_{1:1:6}(6)$                    & {\color[HTML]{6434FC} $p^{(3)}_{1:1:6}(7)$} &                                                                                                   & $p^{(5)}_{1:1:6}(9)$                                                                               &                                                                                                   & $p^{(7)}_{1:1:6}(11)$                &                                      \\ \cline{2-11} 
                         &                                        &                                        & $p^{(0)}_{1:1:6}(5)$                   & $p^{(1)}_{1:1:6}(6)$                    & $p^{(2)}_{1:1:6}(7)$                        & {\color[HTML]{6434FC} $p^{(3)}_{1:1:6}(8)$}                                                       &                                                                                                    & $p^{(5)}_{1:1:6}(10)$                                                                             &                                      &                                      \\ \cline{2-11} 
                         &                                        &                                        &                                        & $p^{(0)}_{1:1:6}(6)$                    & $p^{(1)}_{1:1:6}(7)$                        & $p^{(2)}_{1:1:6}(8)$                                                                              & {\color[HTML]{6434FC} $p^{(3)}_{1:1:6}(9)$}                                                        &                                                                                                   &                                      &                                      \\ \cline{2-11} 
                         &                                        &                                        & {\color[HTML]{FE0000} $m_{2:2:16}(4)$} & {\color[HTML]{FE0000} }                 & {\color[HTML]{FE0000} }                     & {\color[HTML]{FE0000} }                                                                           & {\color[HTML]{FE0000} }                                                                            & {\color[HTML]{FE0000} }                                                                           &                                      &                                      \\ \cline{2-11} 
                         &                                        &                                        &                                        & {\color[HTML]{FE0000} $m_{18:2:24}(4)$} & {\color[HTML]{32CB00} }                     & {\color[HTML]{32CB00} }                                                                           & {\color[HTML]{32CB00} }                                                                            & {\color[HTML]{32CB00} }                                                                           & {\color[HTML]{32CB00} }              & {\color[HTML]{32CB00} }              \\ \cline{2-11} \multirow{-11}{*}{\rotatebox[origin=c]{90}{Relay}}
                         &                                        &                                        &                                        &                                         & {\color[HTML]{963400} $m_{1:2:24}(4)$}      &  \makecell{$p^{(4)}_{1:1:12}(8)$ \\ {\color[HTML]{FE0000}$=m_{2:2:24}(4) $} \\ {\color[HTML]{963400}$+ m_{1:2:24}(4)$}} & \makecell{$p^{(4)}_{1:1:12}(8)$ \\ {\color[HTML]{FE0000}$=m_{2:2:24}(4) $} \\ {\color[HTML]{963400}$+ 2m_{1:2:24}(4)$}} & \makecell{$p^{(4)}_{1:1:12}(8)$ \\ {\color[HTML]{FE0000}$=m_{2:2:24}(4) $} \\ {\color[HTML]{963400}$+3m_{1:2:24}(4)$}} & {\color[HTML]{32CB00} }              & {\color[HTML]{32CB00} }              \\ \cline{2-11} 
 &                                        &                                        &                                        &                                         & {\color[HTML]{32CB00} $m_{2:2:16}(6)$}      & {\color[HTML]{32CB00} $m_{1:2:16}(6)$}                                                            & {\color[HTML]{32CB00} $m_{17:1:24}(6)$}                                                            & {\color[HTML]{32CB00} $p^{(6)}(10)$}                                                              & {\color[HTML]{32CB00} $p^{(6)}(11)$} & {\color[HTML]{32CB00} $p^{(6)}(12)$} \\ \hline
\end{tabular}}

\end{table}

In general, our scheme attempts to transmit each source packet with the maximal possible rate, i.e., $R = \frac{T + 1 - N_1' - N_2}{T + 1 - N_1'}$, where $N_1'$ is the number of erasures observed so far that affect packet $\underline{m}(t)$. As soon as it observes a new erasure, (by updating this $N_1'$) it reduces the rate of transmission of the affected message packets. Further, it also transmits noisy symbols when required, knowing that the noise can always be cancelled at the destination due to the sequential nature of delay-constrained streaming communications. In the following section we present the general code construction. 

\subsection{Code Parameters}

For given parameters $\{N_1,N_2,T\}$, we set message packet, source packet sizes as the following:
\begin{align}
k\triangleq  &\prod_{i=0}^{N_1}{T+1-N_2-i},\label{eq:message_packet_size}\\
n_1 \triangleq &(T+1-N_2)\prod_{i=0}^{N_1-1}{T+1-N_2-i},\label{eq:source_packet_size}\\
n_2 \triangleq &(T+1-N_1)\prod_{i=1}^{N_1}{T+1-N_2-i}\nonumber\\
&+\sum_{l=1}^{N_1}\prod_{i=0,i\neq l}^{N_1}{T+1-N_2-i}.\label{eq:relay_packet_size}
\end{align}
\begin{remark}
    This choice is to ensure that every subcode from relay to destination (which have a rate of the form $(T + 1 - N_2 - i)/(T + 1 - i)$ as mentioned previously) can be met with integer parameters. 
\end{remark}

We represent the message packet $\underline{m}(t)$ as a column vector of the form:
\begin{equation}
\underline{m}(t)\triangleq\begin{bmatrix}
	m_0(t)~m_1(t)~\cdots~m_{k-1}(t)
\end{bmatrix}^T.
\end{equation}

We now proceed to explain the encoding strategy in each link, and show that this encoding strategy results in an achievable rate which is equal to the one stated in Theorem~\ref{theorem:achievable}.
 
\subsection{Source-to-Relay Encoding}\label{sec:source_relay_encoding}

As mentioned previously, the source-to-relay encoding is similar to the previous work on \cite{Silas2019}. The major difference is that we use multiple ``layers'' of the same code. This can be seen in the previous example, where we use 12 layers of a $2/4$ code, that is, we replicate a $2/4$ diagonally-interleaved MDS code twelve times.

\begin{figure*}
	\centering
	\includegraphics[scale=0.55, draft=false]{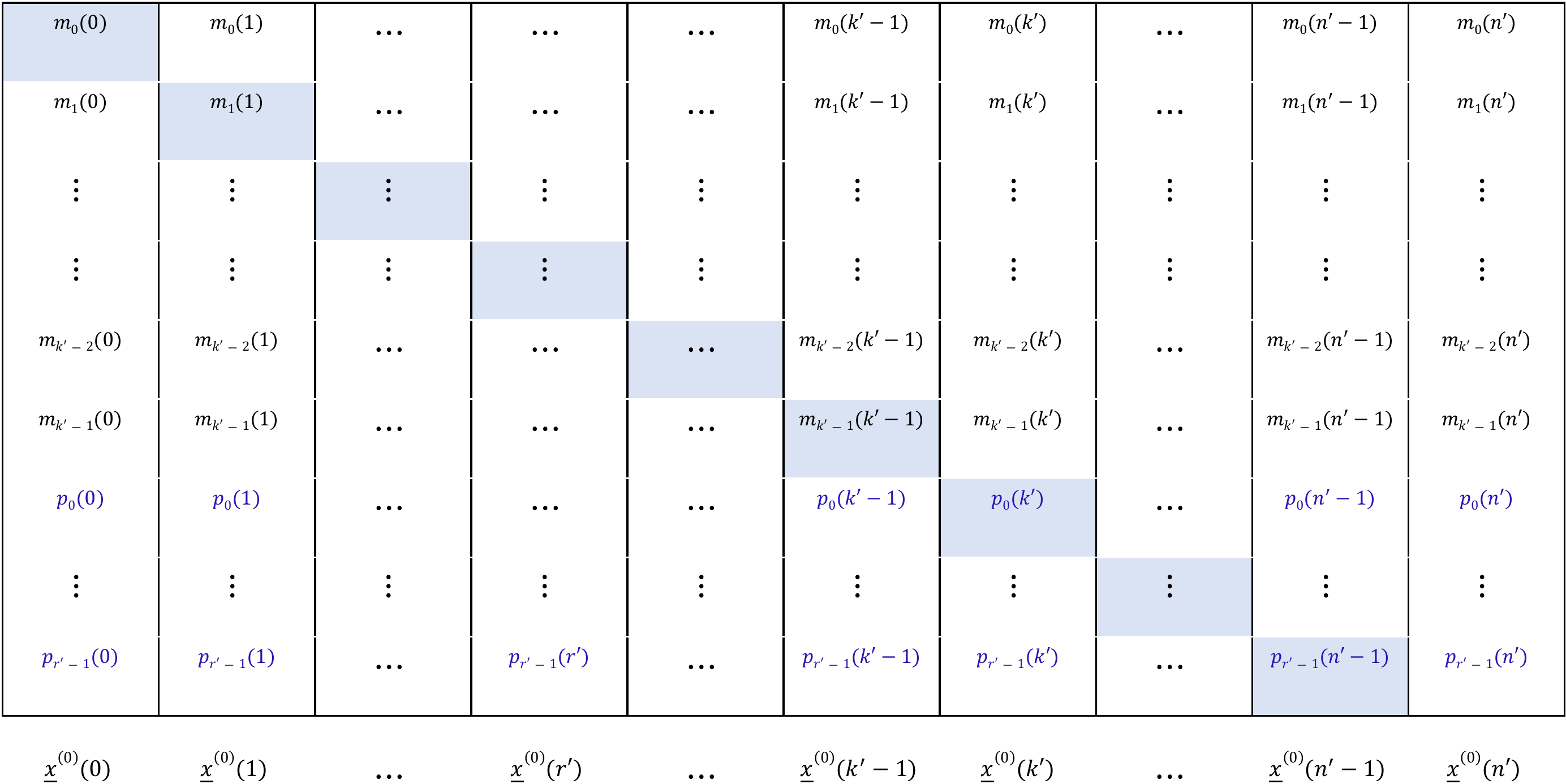}
	\caption{An illustration of diagonal interleaving technique applied by source-side encoder to produce source sub-packets $\{\underline{x}^{(i)}(t)\}_{t\in[0:\infty]}$. We illustrate here the case $i=0$ and the same procedure will be applied for all $i\in[0:\ell'-1]$. Let $k'\triangleq T+1-N_2-N_1,n'\triangleq T+1-N_2,r'\triangleq n'-k'=N_1$ and $\underline{m}^{(0)}(t)\triangleq[m_0(t)\ m_1(t)\ \cdots\ m_{k'-1}(t)]^\top$. Each diagonal is a codeword of a systematic $[n',k']$-MDS code, whose initial $k'$ symbols are message symbols. Coded packet $\underline{x}(t)$ is obtained by vertically stacking $\ell'$ coded sub-packets $\{\underline{x}^{(i)}(t)\}_{i\in[0:\ell'-1]}$.}
	\label{fig:diag_interleaving_source_relay}
\end{figure*}

In general, we use $\ell' \triangleq \prod_{i = 0}^{N_1 - 1} (T + 1 - N_2 - i)$ layers of diagonally-interleaved $[n', k']$-MDS codes with parameters $k' = T + 1 - N_1 - N_2 $ and $n' = T + 1 - N_2$. The construction of the diagonally-interleaved MDS code can be found in, e.g., \cite{Silas2019}. Our code can be constructed by simply ``appending'' the codewords from each layer together. Then, we can write $n_1 = \ell' n'$. The encoding scheme is illustrated in Fig.~\ref{fig:diag_interleaving_source_relay}.

Now, we make a major observation about such codes. Using Lemma~3 from \cite{Silas2019}, we know that, if $\underline{x}(i)$ has been erased, then $\ell'$ symbols of $\underline{m}(i)$ can be recovered at time $i + N_1$, another $\ell'$ symbols can be recovered at time $i + N_1 + 1$, and so on, until the entire message has been recovered. This observation is guaranteed independent of erasure pattern, as long as at most $N_1$ erasures occur. However, considering the erasure pattern, we make a stronger claim about the recovery of symbols and, especially, ``noisy symbols'', which we now define simply as {\it estimates}.
\begin{definition}
    We say $\tilde{m}_j(i)$ is an \textit{estimate} of a source symbol $m_j(i)$ if there exists a function $\Psi_{i, j}$ such that $\Psi_{i, j}( \tilde{m}_j(i), \{\underline{m}(t)\}_{t \in [0:i-1]}) = m_j(i)$.
\end{definition}
That is, $\tilde{m}_j(i)$ is an estimate of $m_j(i)$ if, given past ($i' < i$) message packets, we are able to recover ${m}_j(i)$.
\begin{proposition}\label{prop:recovery}
    Assume the packet $\underline{x}(i)$ is erased. Then, $\ell'$ new estimates of symbols of $\underline{m}(i)$ can be recovered from each subsequent non-erased packet $\underline{x}(i')$, $i' > i$, until estimates of all symbols have been recovered. 
\end{proposition}

To clarify the statement of our proposition, ``new'' estimates mean that the estimates the relay is able to decode from each non-erased packet are always estimates of symbols for which the relay did not have an estimate yet.

To understand this definition and proposition, let us consider the examples given previously. First, consider the example in Table~\ref{tab:spaceder}. %Fig.~\ref{fig:spaceder}. 
In this example, it is straight forward that the relay can recover 12 symbols from $\underline{m}(4)$ at time 5 (from the first non-erased packet after time 4) and another 12 symbols at time 7 (from the second non-erased packet). However, in the example in Table~\ref{tab:burster}, the relay only has access to the so-called \textit{estimates} of 12 symbols of $\underline{m}(5)$ at time 6, since there is still interference from $\underline{m}(4)$. Nonetheless, as can be seen in the example, relaying these estimates is enough, since the destination will have access to previous messages by the deadline of recovery.

\subsection{Relay-to-Destination Encoding}\label{sec:relay_destination_coding}

The relay employs two different encoding mechanisms depending on whether the source packet $\underline{x}(t)$ sent from source is successfully received (non-erased) or not (erased). In each time-$t$, relay transmits a relay packet $\underline{y}(t)$ which is a function of all non-erased source-to-relay source packets within the set $\{\underline{x}(t')\}_{t'\in[0:t]}$. For ease of exposition, we will view each $\underline{y}(t)$ as an unordered set of $n_2$ symbols, rather than a column vector.

\subsubsection{$\underline{x}(t)$ is Non-Erased}\label{sec:x_non_erased}

In this case, since we use systematic encoding from source to relay, the entire message packet $\underline{m}(t)$ is immediately available to the relay at time $t$. The relay will then partition the message packet into $\ell'' \triangleq \prod_{i=1}^{N_1}{(T+1-N_2-i)}$ message sub-packets, denoted as $\{\underline{m}'^{(i)}(t)\}_{i\in[0:\ell''-1]}$. That is, $\underline{m}'^{(i)}(t)$ is the $i$th sub-packet of the message packet $\underline{m}(t)$. From \eqref{eq:message_packet_size} and $\ell''$, it follows that each sub-packet is of size $k'' \triangleq T + 1 - N_2$. The relay will then employ diagonal interleaving using $[n'' \triangleq T + 1, k'']$-MDS codes for each sub-packet, as follows. 

Let $G\triangleq [I_{k''}\ P]$ denote the generator matrix of the $[n'',k'']$-MDS code and let each message sub-packet to be represented as a column vector as follows
\begin{align*}
    \underline{m}'^{(0)}(t)&\triangleq [m'^{(0)}_0(t)\ m'^{(0)}_1(t)\ \cdots\ m'^{(0)}_{k''-1}(t)]^\top= [m_0(t)\ m_1(t)\ \cdots\ m_{k''-1}(t)]^\top \\
    \underline{m}'^{(1)}(t)&\triangleq [m'^{(1)}_0(t)\ m'^{(1)}_1(t)\ \cdots\ m'^{(1)}_{k''-1}(t)]^\top = [m_{k''}(t)\ m_{k''+1}(t)\ \cdots\ m_{2k''-1}(t)]^\top
\end{align*}
and so on. Let
\begin{equation*}
 [p^{(i)}(t+k'')\ p^{(i)}(t+k''+1)\ \cdots\ p^{(i)}(t+n''-1)]=\underline{m}'^{(i)}(t)^\top P.
\end{equation*}
Then, for all $i\in[0:\ell''-1]$, the relay appends $m'^{(i)}_1(t)\ \cdots\ m'^{(i)}_{k''-1}(t),p^{(i)}(t+k'')\ p^{(i)}(t+k''+1)\ \cdots\ p^{(i)}(t+n''-1)$ to $\underline{y}(t),\underline{y}(t+1),\ldots,y(t+n'-1)\triangleq \underline{y}(t+T)\}$, respectively.

That is, each sub-packet is encoded into a diagonal that goes from $t$ up to $t + T$, where the packets at times $t' \in [t:t+k''-1]$ contain systematic symbols, and the remaining packets at times $t' \in [t+k'' : t + T]$ contain parity symbols. Note that there are exactly $N_2$ parity symbols for each sub-packet.

Thus, each non-erased source packet $\underline{x}(t)$ contributes $\ell''$ symbols to each of the relay packets $\underline{y}(t),\underline{y}(t+1),\ldots,y(t+T)$. Note that this is not a systematic streaming code, as the message packet $\underline{m}(t)$ is not fully included in the packet $\underline{y}(t)$. That is, in the relay-to-destination code, all symbols from each sub-packet belong to the same underlying MDS code, unlike in the source-to-relay code.

It should be easy to see that we are able to recover $\underline{m}(t)$ from any $N_2$ erasures using this coding scheme. The rate used is also intuitive: since no erasures occurred from source to relay, we transmit with the same rate as a point-to-point streaming code with delay constraint $T$ and $N_2$ arbitrary erasures.

\subsubsection{ $\underline{x}(t)$ is Erased}\label{sec:x_erased}

On the other hand, if $\underline{x}(t)$ is erased, then the relay has no information of $\underline{m}(t)$ at time $t$, and the relay will follow a different encoding strategy. Let $C(t; j)$ be a set of code symbols (to be viewed as a column vector) computed by the relay as a function of all non-erased source packets in time slots $[0 : t + j]$. The size of each $C(t; j)$ can vary from 0 up to $\ell'$. Our coding strategy consists of including $C(t; i)$ as a part of $\underline{y}(t + i)$, $i \in [1:T]$. In the following, we discuss (i) how to determine $C(t; j)$, (ii) how we obtain a relay packet size which matches \eqref{eq:relay_packet_size} and (iii) how we can guarantee that $\underline{m}(t)$ is recoverable at the destination at time $t + T$ under any $N_2$ erasures in the relay-to-destination link.

Let $\mathcal{I}_t\triangleq\{t_1,t_2,\ldots,t_{T+1-N_2-N_1}\}$ denote the set containing the first ${T+1-N_2-N_1}$ time slots in $[t+1:t+T]$ during which there are no erasures in the source-to-relay link. From Proposition~\ref{prop:recovery}, we know that, at time $t_j$, the relay has access to $\ell' \cdot j$ estimates of symbols of $\underline{m}(t)$, for all $j \in [1 : T + 1 - N_2 - N_1]$.

We start by describing an overview of our coding scheme, without specifying the sizes of each $C(t; j)$. Consider a ``long'' systematic $[n_{\textrm{long}}, k_{\textrm{long}}]$-MDS code, $k_\text{long}\triangleq k$, that is, an MDS code which encodes all $k$ symbols from the message packet $\underline{m}(t)$ into $n_{\textrm{long}}$ coded symbols. The parameter $n_{\textrm{long}}$, as we will explain now, depends on the erasure pattern in the source-relay link. Let us start by constructing a length-$n_\text{long}$ row-vector $C(t)^\top$ which is a codeword of this long MDS code.  

In order to construct $C(t)^\top$, the initial $k$ code symbols of $C(t)^\top$ are $k$ estimates of the symbols of $\underline{m}(t)$. More specifically, the first $\ell'$ code symbols of $C(t)^\top$ are the $\ell'$ estimates of $\underline{m}(t)$ determined by relay at time $t_1$ (i.e., from the first non-erased source packet), the next $\ell'$ code symbols are the $\ell'$ estimates determined at time $t_2$ (second non-erased source packet) and so on, up to the first $k$ symbols of $C(t)^\top$. On the other hand, the final $n_\text{long}-k$ code symbols of $C(t)^\top$ are MDS parity symbols obtained as a function of the initial $k$ code symbols of $C(t)^\top$.

Now, let us discuss how to obtain $C(t, j)$ from $C(t)^\top$. Let us define $\alpha_{t, i}$ as the size of $C(t; i)$, and let our codeword be written as $C(t)^\top\triangleq[C(t;1)^\top\ C(t;2)^\top\ \cdots\ C(t;T)^\top]$. Then, from definition, we should have $n_{\textrm{long}} = \sum_{i\in[1:T]}\alpha_{t,i}$. We now discuss how to determine $\{\alpha_{t,i}\}$. 

Consider time slots $[t: t + T]$. 
From assumption, $\underline{x}(t)$ is erased and there can be at most $N_1-1$ more erasures in time slots $[t+1:t+T]$ (in the source-to-relay link). 
%For $j\in[1:N_1-1]$, let $t+v_j$ denote the $j$-th time slot within $[t+1:t+T]$ where there is an erasure in the link from source to relay. 
%For simplicity in notation, if there are $N_e < N_1 - 1$ erasures in this window, we set $v_{j'}\triangleq T+1$, $j'\in [N_e+1:N_1-1]$, that is, we make the analysis assuming that the remaining erasures occur at time $T+1$. 
%Also, for notation purposes, let us denote $v_0 \triangleq 1$ and $v_{N_1} \triangleq T$. 
Let us denote by $\kappa_t(t + i)$ the cumulative number of estimations of message symbols of $\underline{m}(t)$ available to the relay at time $t + i$. Then, $\alpha_{t, i}$ is obtained as described in Algorithm~\ref{alg:sizecodewords}. Note that for the reminder of this section, $j(i)$ has a special meaning as described in Algorithm~\ref{alg:sizecodewords}, and should not be confused with the generic running index used in notation previously.

\begin{algorithm}
    \caption{Computation of $\alpha_{t, i}$ for $i \in [1:T]$} \label{alg:sizecodewords}
    \begin{algorithmic}
        \STATE $i \gets 1$
        \WHILE{$i \leq T$}
            \STATE $j(i) \gets$ number of erasures in time slots $[t + 1 : t + i - 1]$.
            \STATE $\ell_{j(i)} \gets \frac{k}{T - N_2 - j(i)}$
            \IF{ $\underline{x}(t + i)$ is not erased \OR $\kappa_t(t + i) = k$}
                \STATE $\alpha_{t, i} \gets \ell_{j(i)}$
            \ELSE
                \STATE $\alpha_{t, i} \gets \min \left\{ \ell_{j(i)}, \kappa_t(t + i) - \sum_{a \in [1:i-1]} \alpha_{t, a}\right\} $
            \ENDIF
            \STATE $i \gets i + 1$
        \ENDWHILE
    \end{algorithmic}
\end{algorithm}

To illustrate, let us analyze the code used for the transmission of packet $\underline{m}(4)$ in the example. In Table~\ref{tab:burster}, we start with $i = 1$, that is, we shall analyze how many symbols should be transmitted at time 5, i.e., $\alpha_{4, 1}$. Following our algorithm, $\underline{x}(4 + 1)$ is erased. So far, we have not recovered any symbols from $\underline{m}(4)$, therefore, $\kappa_4(5) = 0$, and we have $\alpha_{4, 1} = 0$. This can be seen from the fact that we do not transmit any symbols from $\underline{m}(4)$ at time 5 in the example. Afterwards, for $i = 2$, we are able to recover $\ell' = 12$ symbols, so we have $\kappa_4(6) = 12$. Following the algorithm, since $\underline{x}(6)$ is not erased, we have $\alpha_{4, 2} = \frac{24}{6 - 3 - 1} = 12$. The same goes for the remaining of the transmission. 

On the other hand, in Table~\ref{tab:spaceder}, we have $\alpha_{4, 1} = \frac{24}{6 - 3 - 0} = 8$, and we have $\kappa_4(5) = 12$, since we recovered 12 symbols. Then, at time 6, $\underline{x}(6)$ is erased, thus, we transmit the minimum between $\ell_{j(2)} = 8$ and $\kappa_{4}(6) - \alpha_{4, 1} = 4$, i.e., 4 symbols. Finally, we have $\alpha_{4, i} = 12$ for all other packets, since now $\ell_{j(i)} = 12$, as before.

This is just a greedy algorithm such that as many symbols are included in $C(t;i)$ subject to following constraints:
\begin{enumerate}
\item $\alpha_{t,i}\leq \ell_{j(i)} \leq \ell'$,
\item $C(t;i)$ is a function of message symbol estimates of $\underline{m}(t)$ obtained by relay in non-erased time slots among $[t+1:t+i]$,
\item  $C(t)^\top\triangleq[C(t;1)^\top\ C(t;2)^\top\ \cdots\ C(t;T)^\top]$ is a codeword of a systematic $[n_\text{long},k_\text{long}]$-MDS code. Initial $k$ code symbols  $k$ message symbol estimates of $\underline{m}(t)$.
\end{enumerate}

Note that, because of item 1, we are guaranteed to always have enough symbols to transmit because, if $\underline{x}(t+i)$ is not erased, then we recover at least as many symbols as we transmit, and if $\underline{x}(t+i)$ is erased, then we transmit the minimum between how many symbols we have available and $\ell_{j(i)}$, which is from definition at most the number of symbols we have available.

\subsubsection{Worst-Case Length of Relay Packets} \label{sec:worstcase}

We now wish to show that, using our code construction, the maximum packet length is at most the one described in \eqref{eq:relay_packet_size}. For consistency, all packets which would have a smaller packet length than the maximum are zero-padded in order to keep a constant packet length as defined in the problem statement.

First, let us note that, if $\underline{x}(t')$ is not erased, it appends exactly $\ell''$ symbols to each relay packet in time slots from $t'$ up to $t' + T$, where recall that $\ell'' = \frac{k}{T+1-N_2}$. Thus, at some arbitrary time $t$, each non-erased source packet $\underline{x}(t')$, $t' \in [t - T : t]$ contributes $\ell''$ symbols to $\underline{y}(t)$. 

Now, let us assume that there are $i \leq N_1$ erasures at time slots $\{\tau_1,\tau_2,\ldots,\tau_i\}\subseteq [t-T:t]$, where $\tau_1 < \tau_2 < \cdots < \tau_i$. Now, note that from time $\tau_{i'}+1$ up to time $t$, $i - i'$ erasures have occurred, from definition. Recall also that $\alpha_{t, i} \leq \ell_{j(i)}$ in Algorithm~\ref{alg:sizecodewords}. Finally, note that $\ell_0 < \ell_1 < \cdots < \ell_{N_1 - 1}$. From these properties, we have that 
\begin{align*}
    \alpha_{\tau_{i'}, t - \tau_{i'}} \leq \ell_{i - i'}, i' \in [1:i] 
\end{align*}
where $\alpha_{\tau_{i'}, t - \tau_{i'}}$ is the number of symbols appended to $\underline{y}(t)$ due to the erasure in the link from source to relay at time $\tau_{i'}$. 

Therefore, the packet length of $\underline{y}(t)$ is at most
\begin{align}
    \tilde{n} &\leq \underbrace{(T + 1 - i) \ell''}_{\text{Contribution of non-erased packets}} +
    \underbrace{ \sum_{i' \in [0 : i - 1]} \ell_{i'}  }_{\text{Contribution of erased packets}} \\
    &= {(T + 1 - i) \ell''} +  \sum_{{i'} \in [0 : i - 1]} \frac{k}{T - N_2 - i'}.
\end{align}
Now, note that $\ell'' < \ell_0$, thus, $i = N_1$ maximizes this packet length and is therefore the worst case. We then have that $\tilde{n} \leq (T + 1 - N_1) \frac{k}{T + 1 - N_2} + \sum_{i' = 0}^{N_1 - 1} \frac{k}{T - N_2 - i'}$, which matches \eqref{eq:relay_packet_size}.

Finally, we should note that, since the relay changes its coding strategy according to the erasure pattern observed in the first link, this erasure pattern must also be relayed to the destination, so it knows how to decode. A naive solution is to, at time $t$, transmit the erasure pattern observed from time $t - T$ up to $t$, which is a binary sequence of length $T + 1$, and does not depend on the packet size. Thus, by making the packet size go to infinity, the rate approaches
\begin{equation*}
    R_2 = \frac{T + 1 - N_2}{T + 1 + \sum_{i = 0}^{N_1 - 1} \frac{N_1 - i}{T + 1 - N_2 - (N_1 - i)}}.
\end{equation*}

\subsubsection{Recoverability of $\underline{m}(t)$ at Destination by Time-$(t+T)$} \label{sec:recover}

We have shown that our code construction achieves the desired rate, however, it remains to show that our code is $(N_1, N_2, T)$-achievable, and therefore so is the proposed rate.

Note that Proposition~\ref{prop:recovery} shows that the relay is able to recover the estimates of $\underline{m}(t)$. Now, it suffices to show that, with the proposed code construction, the destination has access to enough estimates to recover $\underline{m}(t)$ entirely.
\begin{proposition}
    Using our coding scheme, if there are at most $N_2$ erasures from relay to destination, the destination is able to recover an estimate $\tilde{\underline{m}}(t)$ of $\underline{m}(t)$ at time $t + T$.
\end{proposition}

\begin{IEEEproof}
    Note that, from relay to destination, all message packets are transmitted within the same MDS code. We split the analysis in two cases, depending on whether $\underline{x}(t)$ has been erased or not. If $\underline{x}(t)$ has not been erased, then each message sub-packet is encoded using a $[T + 1, T + 1 - N_2]$-MDS code, and at most $N_2$ erasures may occur, therefore, it is straight forward that all $(T + 1 - N_2)\ell'' = k$ symbols can be recovered, since each layer recovers its own $(T + 1 - N_2)$ symbols and there are $\ell''$ layers.
    
    On the other hand, if $\underline{x}(t)$ has been erased, we again split the analysis in two cases. In the first case, let us assume $\alpha_{t, i} = \ell_{j(i)}$ for every $i$, that is, we always have enough symbols to transmit the maximum we wish to. In this case, $\alpha_{t, i} \geq \frac{k}{T - N_2}$ for all $i \in [1:T]$. Since there are at most $N_2$ erasures, there are at least $T - N_2$ available packets, thus we can recover all $k$ message symbols.
    
    Finally, the last case occurs when $\underline{x}(t)$ has been erased, but $\alpha_{t, i} < \ell_{j(i)}$ for some $i$. In this case, let us denote by $i^*$ the largest $i$ such that this condition holds. Let us state some easily verifiable facts about $i^*$:
    \begin{enumerate}
        \item $\underline{x}(t + i^*)$ has been erased, as otherwise we would have $\alpha_{t, i^*} = \ell_{j(i^*)}$.
        \item There are $j^* \triangleq j(i^*) + 1$ erasures in the source-to-relay link in time slots $[t+1 : t + i^*]$. %$i^* = v_{j^*}$ for some $j^* \in [1 : N_1 - 1]$.
        \item $\kappa_t(t + i^*) < k$, as otherwise we would have $\alpha_{t, i^*} = \ell_{j(i^*)}$.
        \item $\sum_{i = 1}^{i^*} \alpha_{t, i} = \kappa_t(t + i^*)$.%, as otherwise we would have $\alpha_{t, i^*} = \ell_{j(i^*)}$. 
    \end{enumerate}
    Then, for $i \in [i^* + 1 : T]$, we have
    \begin{equation}
        \alpha_{t, i} \geq \frac{k}{T - N_2 - j^*} \label{eq:cond1}.
    \end{equation}
    This follows from the fact that, for $i > i^*$, we have $\alpha_{t, i} = \ell_{j(i)}$ from the definition of $i^*$ and $j(i) \geq j^*$ from definition of $j(i)$ and $j^*$.
    
    Also, for $i \in [1: i^*]$, we have 
    \begin{equation}
        \alpha_{t, i} \leq \frac{k}{T - N_2 - (j^* - 1)}. \label{eq:cond2}
    \end{equation}
    This again follows from the fact that, for $i \leq i^*$, $j(i) \leq j^* - 1$ (recall that $j(i)$ only counts erasures up to time $t + i - 1$). 
    
    Also, note the following: $i^* - j^*$ is exactly the number of non-erased source packets from time $t+1$ up to $t+i^*$, since $i^*$ represents the number of transmitted packets and $j^*$ the number of erased packets. Furthermore, we have
    \begin{align*}
        \ell' (T + 1 - N_1 - N_2) &\overset{(a)}{=} k \\
        &\overset{(b)}{>} \kappa_t(t + i^*) \overset{(c)}{=} (i^* - j^*) \ell'
    \end{align*}
    where $(a)$ follows from definition \eqref{eq:message_packet_size}; $(b)$ follows from item 3 above; and $(c)$ follows from the fact that there are $i^* - j^*$ non-erased packets and the relay recovers $\ell'$ estimates from each non-erased packet as seen in Proposition~\ref{prop:recovery}. This implies 
    \begin{equation}
        T - N_2 - i^* > N_1 - 1 - j^* \geq 0 \label{eq:cond3}
    \end{equation}
    where the first inequality is simply from rewriting the previous inequality and the second comes from the fact that $j^* \leq N_1 - 1$, since at most $N_1 - 1$ erasures may occur from time $t+1$ up to $t+T$ (recall that $\underline{x}(t)$ has been erased). 
    
    For simplicity, let us define as $\mathcal{I}$ the indices $i$ such that $t + i$ has not been erased in the link from relay to destination. Then, it follows that
    \begin{align}
        \sum_{i \in \mathcal{I}} \alpha_{t, i} &\overset{(a)}{\geq} \sum_{i = 1}^{i^*} \alpha_{t, i} + (T - i^* - N_2) \frac{k}{T - N_2 - j^*} \\
        &\overset{(b)}{=} (i^* - j^*) \frac{k}{T + 1 - N_1 - N_2} + (T - i^* - N_2) \frac{k}{T - N_2 - j^*} \\
        &\overset{(c)}{\geq} (i^* - j^*) \frac{k}{T - N_2 - j^*} + (T - i^* - N_2) \frac{k}{T - N_2 - j^*} \\
        &= k
    \end{align}
    where $(a)$ follows from the fact that, in the worst case, from conditions \eqref{eq:cond1} and \eqref{eq:cond2}, all $i \in [1:i^*]$ are part of $\mathcal{I}$ (otherwise we get more symbols, that is, if $i' > i^*$ and $i'' \leq i^*$, then $\alpha_{t, i'} > \alpha_{t, i''}$), and that the remaining can all be bounded by \eqref{eq:cond1}; $(b)$ follows from the fact that $\sum_{i=1}^{i^*} \alpha_{t, i} = \kappa_t(t + i^*) = (i^* - j^*) \ell' = (i^* - j^*) \frac{k}{T + 1 - N_1 - N_2}$, from item 4 above and the fact that, for each non-erased source packet, the relay recovers $\ell'$ estimates; and $(c)$ comes from the fact that, again, $j^* \leq N_1 - 1$. The last step results from trivial arithmetic. 
    
    Then, owing to the use of the long $[n_\text{long},k_\text{long}=k]$-MDS code, since at least $k$ code symbols are available, then all $k$ estimates of message symbols of $\underline{m}(t)$ can be recovered by the delay constraint at time $(t+T)$. 
    
    Finally, note that messages $\underline{m}(t')$, $t' < 0$ are available from definition to the destination. Therefore, at time $T$, since the destination is able to recover an estimate of $\underline{m}(0)$, it is also able to recover the message packet itself. Then, at time $T + 1$, since it already has access to $\underline{m}(0)$, the estimate of $\underline{m}(1)$ is enough to recover the message packet as well, and so on. Thus, from this induction argument, it is easy to see that the estimates are enough for the destination to recover the message packets by the deadline.
\end{IEEEproof}

Finally, we propose a naive way to inform the destination about the erasure pattern that has been observed: at every time instant $t$, the relay forwards the observed erasure pattern from time $t - T$ up to time $t$. This is a binary sequence of length $T+1$, thus it can be represented by $\lceil (T + 1) \log_q 2 \rceil$ symbols. Further, it does not depend on the packet size, thus, we can make this overhead negligible by increasing the packet sizes. This can be easily achieved by multiplexing together $c$ copies of our code\footnote{For example, for $T = 5$ and $q = 2^8$, this overhead is a one-byte header, which is negligible in a 256 bytes packet.}.

Finally, since we have presented an $(N_1, N_2, T)$-achievable code with rate as described in Theorem~\ref{theorem:achievable}, the proof of the theorem is complete.

\section{Upper Bound}

\begin{table}
\centering \caption{A valid erasure pattern for $N_1=1$, $N_2=2$ and $T = 4$.} \label{tab:erasuretoyupper}
\begin{tabular}{|c|l|l|l|l|l|l|l|l|}
\hline
Time              & $t$                      & $t+1$ & $t+2$                    & $t+3$ & $t+4$                    & $t+5$ & $t+6$ & $t+7$                    \\ \cline{2-9} 
Source-Relay      & \cellcolor[HTML]{FE0000} &       &                          &       &                          &       &       & \cellcolor[HTML]{FE0000} \\ \cline{2-9} 
Relay-Destination & \cellcolor[HTML]{FE0000} &       & \cellcolor[HTML]{FE0000} &       & \cellcolor[HTML]{FE0000} &       &       & \cellcolor[HTML]{FE0000} \\ \hline
\end{tabular}
\end{table}

We start the upper bound presentation by showing a toy-case example for which we show that an $(N_1 = 1, N_2 = 2, T = 4)$-achievable code must be able to recover from more than $N_2$ erasures in the second link under some particular conditions. Let us consider the erasure pattern in Table~\ref{tab:erasuretoyupper}. Note that, from time $t$ to $t + T$, there are $N_2 + 1 = 3$ erasures. However, consider the following argument: at time $t + T$, all source packets from time $0$ up to $t - 1$ must have been recovered. Furthermore, since $\underline{x}(t)$ has been erased, and due to causality, packet $\underline{y}(t)$ must be a function only of packets from time $0$ up to $t-1$, as the relay has not yet received any information about packet $\underline{m}(t)$. Therefore, at time $t + T$, that is, at the deadline for recovery of packet $\underline{m}(t)$, the destination must be able to generate $\underline{y}(t)$, thus its erasure should not affect the recovery of $\underline{m}(t)$. This example is formalized in the following entropy equations
\begin{align*}
    H(\underline{m}(t) | \{ \underline{y}(t') \}_{t' = 0}^{t - 1}, \underline{y}(t+1), \underline{y}(t+3)) &\overset{(a)}{=} H(\underline{m}(t) | \{ \underline{y}(t') \}_{t' = 0}^{t - 1}, \underline{y}(t+1), \underline{y}(t+3), \{\underline{m}(t')\}_{t' = 0}^{t - 1}) \\
    &\overset{(b)}{=} H(\underline{m}(t) | \{ \underline{y}(t') \}_{t' = 0}^{t}, \underline{y}(t+1), \underline{y}(t+3), \{\underline{m}(t')\}_{t' = 0}^{t - 1}) \\&\overset{(c)}{=} 0
\end{align*}
where $(a)$ follows from the fact that, at time $t + 3$, all packets from time 0 up to $t - 1$ must be recoverable, since, up to this point, at most $N_2$ erasures have occurred in the second link, and at most $N_1$ in the first link; $(b)$ follows from the fact that, since $\underline{x}(t)$ has been erased, $\underline{y}(t)$ must be a function only of message packets up to time $t-1$, due to causality; and $(c)$ follows from the fact that only $\underline{y}(t + 2)$ and $\underline{y}(t + 4)$ are missing, thus it is as if only $N_2 = 2$ erasures have occurred, and therefore $\underline{m}(t)$ must be recoverable. Then, by repeating this erasure pattern with a period of $7$, and simply computing the ratio of non-erased packets to total packets in the second link, we get that $R_2 \leq \frac{4}{7}$. On the other hand, the trivial bound in this scenario would be $R_2 \leq  \frac{T + 1 - N_2}{T + 1} = \frac{3}{5}$. If we consider only time-invariant codes, then the upper bound is given by $\frac{2}{4}$, which is clearly lower than ours. However, it applies only to time-invariant codes, and in fact we can achieve higher than that using our adaptive coding scheme presented in the previous section, which, in this example, achieves $\frac{3}{5.5}$, ignoring the overhead.

The following lemma generalizes this key concept and allow us to make an induction argument in the sequence.

\begin{lemma} \label{lemma:recoverabilityt}
    Assume that, at time $t + T$, the destination has access to message packets \{$\underline{m}(t')\}_{t' = 0}^{t - 1}$. Then, if the following conditions hold, an $(N_1, N_2, T)$-achievable code must be able to recover $\underline{m}(t)$.
    \begin{align}
        \sum_{t' = t}^{t + T} e_{S, t'} &\leq N_1 \label{eq:condition1} \\
        \sum_{t' = t}^{t + T} e_{R, t'} &\leq N_2, \textrm{~if~} e_{S, t} = 0 \\
        \sum_{t' = t+1}^{t + T} e_{R, t'} &\leq N_2, \textrm{~if~} e_{S, t} = 1. \label{eq:condition2}
    \end{align}
\end{lemma}
\begin{IEEEproof}
    % First, let us formalize, in terms of entropy, what a code being $(N_1, N_2, T)$-achievable means. Let us denote as $\mathcal{X}_{T} = \{ \underline{x}(t') : e_{S, t'} = 0, 0 \leq t' \leq t + T\}$, that is, the subset of non-erased channel packets transmitted from source to relay. Since $\underline{y}_D(t)$ only depends on $\underline{m}(t)$ through $\underline{x}(t)$, it is clear that, if the destination must recover from $N_1$ erasures in the first link, then, for any $\mathcal{X}_T$ such that $|\mathcal{X}_T| \geq t + T + 1 - N_1$, we have
    % \begin{equation}
    %     H( \{\underline{m}(t') \}_{t' = 0}^{t} | \mathcal{X}_T) = 0.
    % \end{equation}
    % Further, let us denote as $\mathcal{Y}_{T} = \{ \underline{y}(t') : e_{R, t'} = 0, 0 \leq t' \leq t + T \}$, that is, the subset of non-erased channel packets transmitted from relay to destination. From the same argument as before, if $|\mathcal{Y}_T| \geq t + T + 1 - N_2$, we have
    % \begin{equation}
    %     H( \{\underline{m}(t') \}_{t' = 0}^{t} | \mathcal{Y}_T) = 0.
    % \end{equation}
    % Finally, we also note that $H( \mathcal{Y}_T | \mathcal{X}_T) = 0$, since the non-erased symbols from relay to destination are a function of the non-erased symbols from source to relay.
    %
    Let us split the proof in two cases. First, if $e_{S, t} = 0$, the proof follows immediately
    \begin{align}
        H(\underline{m}(t) | \{ \underline{m}(t')\}_{t' = 0}^{t - 1}, \{\underline{y}_D(t')\}_{t' = t}^{t + T}) &= H(\underline{m}(t) | \{ \underline{m}(t')\}_{t' = 0}^{t - 1}, \{\underline{y}_D(t')\}_{t' = t}^{t + T}, \{ \underline{y}(t'), \underline{x}(t')\}_{t' = 0}^{t - 1}) \\
        & = 0
    \end{align}
    since, from assumption, the code is $(N_1, N_2, T)$-achievable, therefore, it must be able to recover from any $N_1$ erasures in the first link and $N_2$ erasures in the second link. In particular, since the destination has access to all previous message packets, it is able to generate all channel packets from time $0$ up to $t - 1$, independent of the erasures that have occurred in the past. Therefore, it is ``as if'' only the erasures from time $t$ up to $t + T$ have occurred. 
    
    Now, let us consider the more interesting case, which is when $e_{S, t} = 1$. In this case, we have
    \begin{align}
        H(\underline{m}(t) | \{ \underline{m}(t')\}_{t' = 0}^{t - 1}, \{\underline{y}_D(t')\}_{t' = t+1}^{t + T}) &= H(\underline{m}(t) | \{ \underline{m}(t')\}_{t' = 0}^{t - 1}, \{\underline{y}_D(t')\}_{t' = t+1}^{t + T}, \{ \underline{y}(t')\}_{t' = 0}^{t}, \underline{x}(t')\}_{t' = 0}^{t-1}) \\
        & = 0
    \end{align}
    where, again, since the destination has access to previous message packets, it is able to generate all the channel packets from source to relay $\{\underline{x}(t')\}_{t'=0}^{t-1}$. However, since $e_{S, t} = 1$, we have that $\underline{y}(t)$ must also be a function of $\{\underline{x}(t')\}_{t'=0}^{t-1}$, thus the destination can also generate the channel packet $\underline{y}(t)$. Then, again, it is ``as if'' only the erasures from time $t + 1$ up to $t + T$ have occurred from relay to destination, and, since there are at most $N_2$ erasures in this window, packet $\underline{m}(t)$ must be recoverable. 
\end{IEEEproof}

We now apply this Lemma in an induction argument in order to state that not only we must be able to recover from a sliding window, as shown in e.g. \cite{martinian2004burst} and recalled in Remark~\ref{remark:recover}, we must be able to recover from a sliding window that, sometimes, allows for more than $N_2$ erasures in the second link.

\begin{lemma}\label{lemma:upperboundR2}
    If a code is $(N_1, N_2, T)$-achievable, then it must be able to correct any erasure patterns for which the following holds:
    \begin{align}
        \sum_{t' = t}^{t + T} e_{S, t'} &\leq N_1, \forall t \in \mathbb{Z}_+ \\
        \sum_{t' = t}^{t + T} e_{R, t'} &\leq N_2, \forall t \in \{t\in \mathbb{Z}_+: e_{S, t} = 0\} \\
        \sum_{t' = t+1}^{t + T} e_{R, t'} &\leq N_2, \forall t \in \{t\in \mathbb{Z}_+: e_{S, t} = 1\} \label{eq:newcond2}
    \end{align}
\end{lemma}
\begin{remark}
    These conditions lead to a tighter upper bound than the trivial bound that only considers the second link, which is represented by the condition $\sum_{t' = t}^{t + T} e_{R, t'} \leq N_2, \forall t \in \mathbb{Z}_+$. In particular, \eqref{eq:newcond2} allows for more erasures in a window than the trivial bound allows, thus, there are less available non-erased packets and the rate must be lower. In other words, the erasure sequence used in the trivial bound meets the conditions of Lemma~\ref{lemma:upperboundR2}, and allowing for more possible erasure sequences can only improve the upper bound.
\end{remark}
\begin{IEEEproof}
    This follows directly from an induction argument using Lemma~\ref{lemma:recoverabilityt}. Note that, from definition, at time $T$, the destination has access to message packets $\underline{m}(t')$, $t' < 0$. Therefore, if conditions \eqref{eq:condition1}-\eqref{eq:condition2} hold for $t = 0$, packet $\underline{m}(0)$ is recoverable. Then, at time $T + 1$, the destination has access to $\underline{m}(t)$, and if the conditions hold for $t = 1$, it can also recover $\underline{m}(1)$, and so on.
    
    Formally, assume that, at time $\tau + T$, the destination has access to all message packets $\underline{m}(t')$, $t' < \tau$. Then, under the assumptions of this Lemma and applying Lemma~\ref{lemma:recoverabilityt}, at time $\tau + T + 1$, the destination will have access to all message packets $\underline{m}(t')$, $t' \leq \tau$. That is, if, at any time instant, all previous message packets are known to the destination, then, under our assumptions and using Lemma~\ref{lemma:recoverabilityt}, all following message packets will also be recoverable by the destination. Finally, remember that, by assumption from the model, $\underline{m}(t')$, $t' < 0$ is known to the destination. Therefore, under the assumptions of this Lemma, all message packets are recoverable by the destination.
\end{IEEEproof}

This Lemma provides us a framework to try and find an optimal erasure pattern pair which fulfills the desired constraints. 
\begin{proposition}
    An upper bound to the rate in the second link can be found by solving the following optimization problem 
    \begin{align}
    R_2^* \leq \lim_{\tau \to \infty} \quad &\min_{e_S^\infty, e_R^\infty} \frac{1}{\tau} \sum_{t' = 0}^{\tau + T} (1 -  e_{R, t'}) \nonumber\\
    \textrm{s.t.} \quad &  \sum_{t' = t}^{t + T} e_{S, t'} \leq N_1, \forall t \in \mathbb{Z}_+ \nonumber \\
        &\sum_{t' = t}^{t + T} e_{R, t'} \leq N_2, \forall t \in \{t\in \mathbb{Z}_+: e_{S, t} = 0\}  \nonumber\\
        &\sum_{t' = t+1}^{t + T} e_{R, t'} \leq N_2, \forall t \in \{t\in \mathbb{Z}_+: e_{S, t} = 1\} \label{eq:optimizationproblem}
    \end{align}
    where the optimization is over the valid erasure sequences that satisfy Lemma~\ref{lemma:upperboundR2}.
\end{proposition}
\begin{IEEEproof}
    We have shown, in Lemma~\ref{lemma:upperboundR2}, that if a code is $(N_1, N_2, T)$-achievable, it must be able to recover from an erasure sequence that meets the conditions in Lemma~\ref{lemma:upperboundR2}. Given a valid erasure sequence, it is easy to see that the rate must be below the ratio of non-erased packets (using the given erasure sequence) to total packets. This holds for any window of length $\tau$, and it gets tighter as $\tau \to \infty$, as the difference between $\tau$ (total source packets) and $\tau + T$ (total encoded packets) becomes negligible. 
    
    That is, the optimization is simply finding the valid erasure sequence that meets the conditions given by Lemma~\ref{lemma:upperboundR2}.
\end{IEEEproof}
In our example, an erasure sequence that satisfies Lemma~\ref{lemma:upperboundR2} can be found by repeating the pattern in Table~\ref{tab:erasuretoyupper} with period 7. This can be seen by the fact that, when an erasure occurs in the first link (e.g. at time $t$), the constraint is that the number of erasures from time $t+1$ up to $t + 4$ is less than or equal to $N_2$, which holds. In other windows, where there are no erasures in the first link, the constraint is that the number of erasures is at most $N_2$, which again holds. Then, an upper bound is simply given by the ratio between non-erased packets (4) to total packets (7) for this specific erasure pattern. However, we do not claim that the erasure pattern presented in Table~\ref{tab:erasuretoyupper} is optimal, instead, it is one valid erasure sequence which results into one valid bound.

However, solving this optimization problem is not trivial. Instead, we propose a heuristic algorithm for the adversary, which attempts to maximize the number of erasures it introduces in the relay-destination channel. The algorithm is described in Algorithm~\ref{alg:heuristic}.
\begin{algorithm}
    \caption{Adversary Heuristic} \label{alg:heuristic}
    \begin{algorithmic}
        \FOR{ $t = 0$, $t$++, $t \leq \tau$}
            \STATE num\_erasures\_R$(t) \gets \sum_{t' = t}^{t + T} e_{R, t'}$
            \IF{$e_{R, t} == 1$}
                \STATE  num\_erasuresS$(t) \gets \sum_{t' = t - T}^{t} e_{S, t'}$
                \IF{num\_erasures\_R$(t) == N_2$ \AND num\_erasures\_S$(t) < N_1$}
                    \STATE $e_{S, t} \gets 1$
                \ENDIF
            \ENDIF
            
            \IF{($e_{S, t} == 1$ \AND num\_erasures\_R$(t) \leq N_2$) \OR  (num\_erasures\_R$(t) < N_2$)}
                \STATE $e_{R, t + T} \gets 1$
            \ENDIF
        \ENDFOR
    \end{algorithmic}
\end{algorithm}

The idea for the algorithm is simple: we introduce erasures in the second link in a greedily manner, always in the end of the window (i.e., at time $t + T$). This ensures that introducing this erasure will not make any previous window invalid, that is, it will not increase the number of erasures in any previous window to more than it is allowed. Furthermore, under our constrained framework, introducing erasures in the first link only improves the upper bound by allowing one extra erasure in the second link for the same time instant. For that reason, we only introduce erasures in the first link at times when there is already an erasure in the second link \textbf{and} that being able to introduce one extra erasure helps, that is, when the number of erasures in the window is already $N_2$. 

Further, we note that the source is not able to adapt to any erasure pattern, thus the cut-set-like bound from \cite{Silas2019} for the first link holds even in the adaptive case. That is, as in \cite{Silas2019}, we can upper bound the capacity by analyzing the first link and noting that the relay must recover every message packet with delay $T - N_2$, otherwise a burst of $N_2$ erasures in the second link makes that information impossible to recover. Thus, as in the prior works, we can bound $R_1 \leq \frac{T + 1 - N_1 - N_2}{T + 1 - N_2}$. Finally, this allows us to present the following upper bound.

\begin{theorem}
    The $(N_1, N_2, T)$-capacity $C_{N_1, N_2}$ is upper bounded by
    \begin{align*}
        C_{N_1, N_2} &\leq \min(R_1, R_2) \\
        R_1^* &\leq \frac{T + 1 - N_1 - N_2}{T + 1 - N_2} \\
        R_2^* &\leq \eqref{eq:optimizationproblem}
    \end{align*}
\end{theorem}
\begin{IEEEproof}
    This follows immediately from the cut-set-like bound in \cite{Silas2019} and Lemma~\ref{lemma:upperboundR2}.
\end{IEEEproof}

\section{Results}

To validate our construction, we consider multiple settings with the parameters $(N_1, N_2, T)$ drawn randomly with the following conditions:
\begin{itemize}
    \item $10 \geq N_2 > N_1 \geq 1$. The condition $N_2 > N_1$ is so that the second link is the bottleneck. Otherwise, both our coding scheme and the non-adaptive coding scheme from \cite{Silas2019} are optimal.
    \item $N_1 + N_2 + 10 \geq T \geq N_1 + N_2$. The second condition is so that the capacity is not trivial (i.e. not zero).
\end{itemize}
With the range of parameters, we hope to observe most meaningful scenarios: $N_2$ and $N_1$ can be close or fairly separated and the delay constraint $T$ can be tight or loose with respect to the number of erasures.

\begin{figure}
    \centering
    \includegraphics[draft=false, scale=.6]{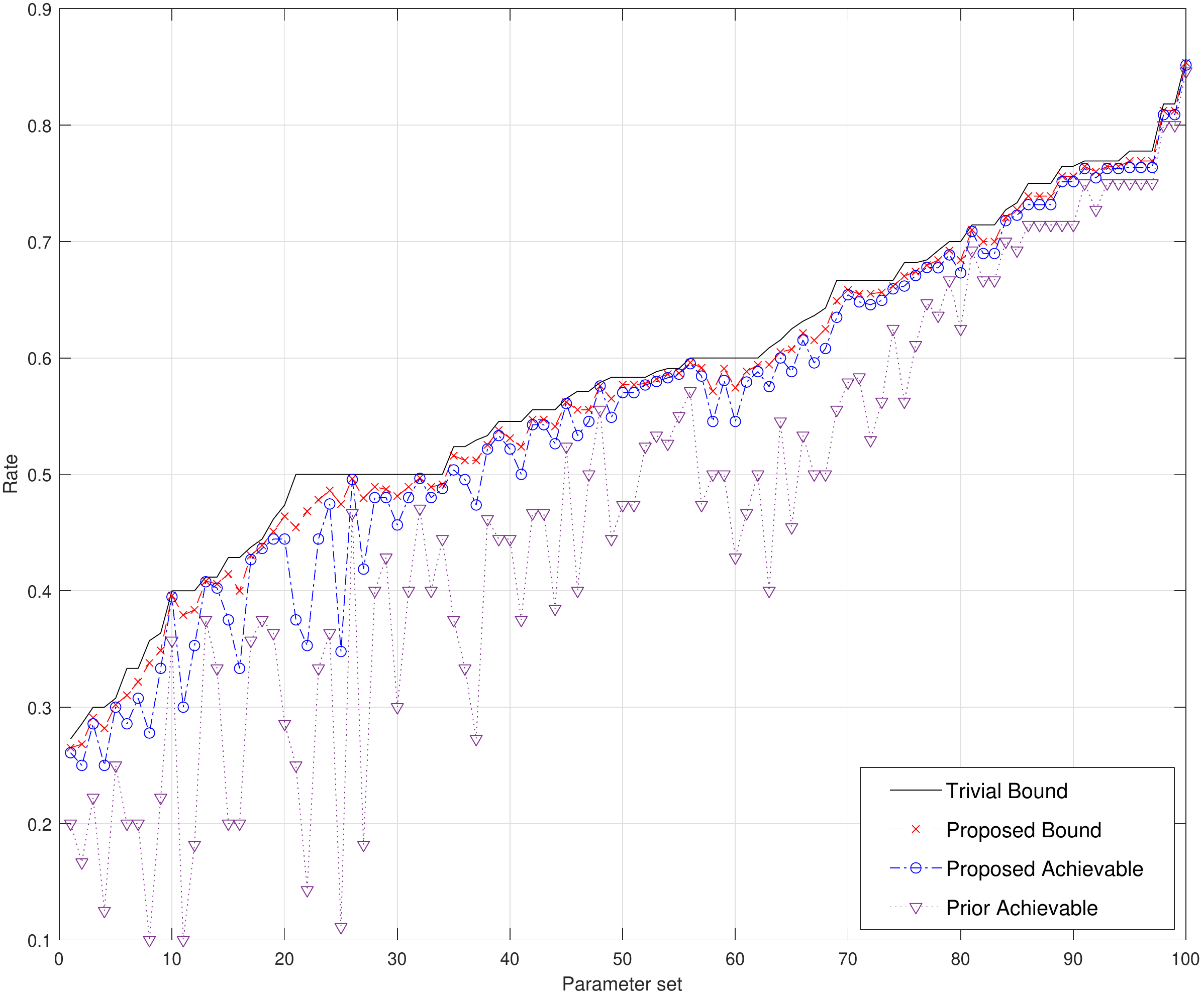}
    \caption{Comparison between prior work and our work in terms of lower and upper bounds on the capacity of streaming codes in the three-node relayed network.}
    \label{fig:results}
\end{figure}

Then, for each randomly chosen set of parameters, we compute a trivial upper bound, which is obtained by reducing the relayed-setting to a point-to-point setting from relay to destination, the achievable rate from \cite{Silas2019}, our upper bound and our achievable rate. Then, for presentation, we sort the results according to the trivial bound, in increasing order, and present it in Fig.~\ref{fig:results}.

It can be seen that our achievable rate is strictly better than the non-adaptive achievable rate, with gains above 100\% for some sets of parameters. We also observed that our upper bound is tighter than the trivial bound in all scenarios, although the difference might be negligible for some setting parameters.

\begin{figure}
    \centering
    \includegraphics[draft=false, trim={4cm, 8.4cm, 4.2cm, 9.1cm}, clip]{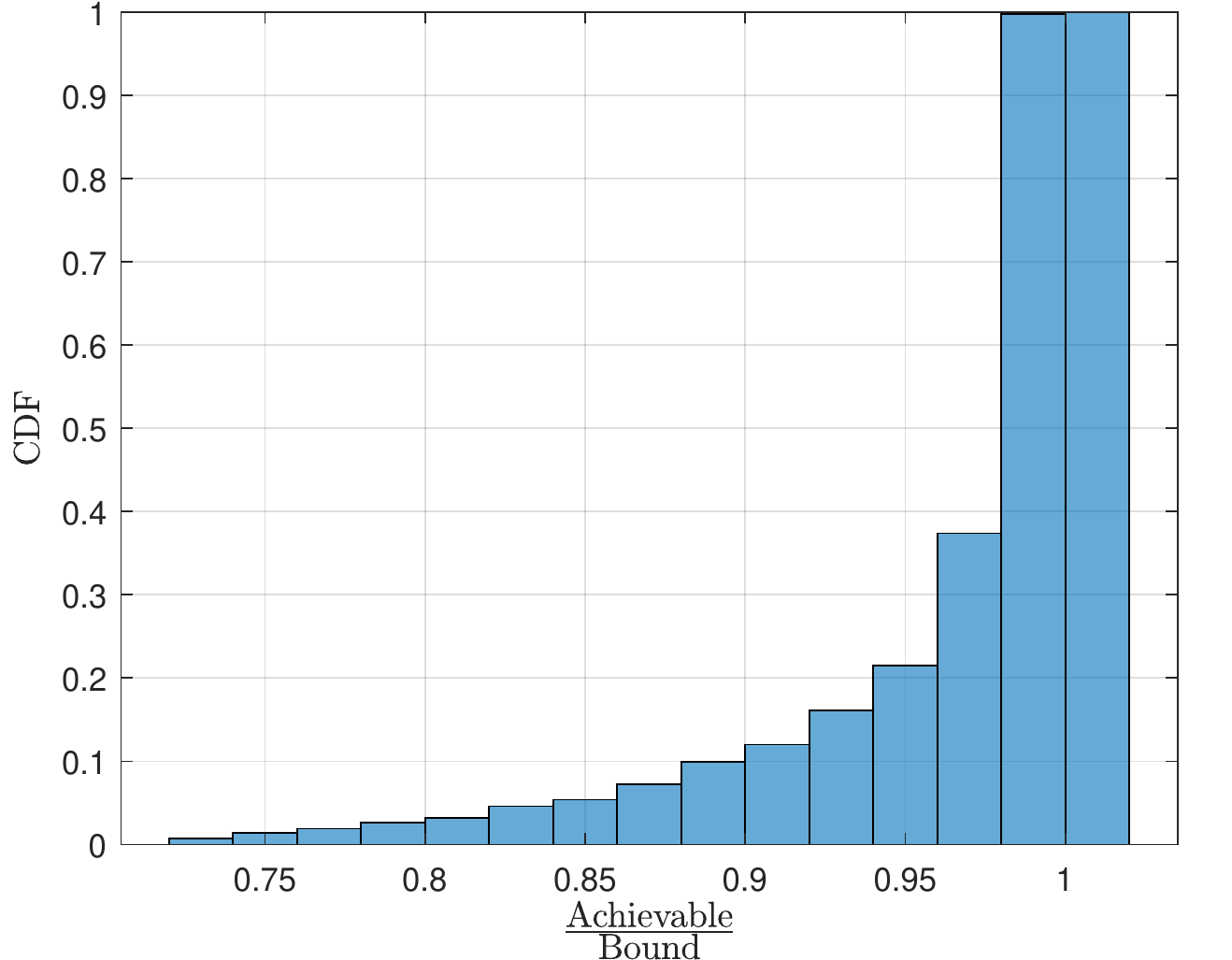}
    \caption{CDF of proposed achievable rate divided by proposed upper bound}
    \label{fig:results2}
\end{figure}

We also present the CDF of the ratio between our achievable rate and our upper bound. As can be seen in Fig.~\ref{fig:results2}, our coding scheme achieves higher than 95\% of the upper bound in more than 80\% of the tested scenarios, and it achieves less than 80\% of the upper bound in less than 3\% of the scenarios.

\section{Conclusion}

In our paper, we presented a novel adaptive scheme for the delay-constrained relay setting, which exploits the fact that the relay can act based on the observed erasure pattern. We also present a novel upper bound technique, which bounds the rate in the second link according to the erasure pattern observed in the first link. We compare both our achievable and upper bound to the prior work on them, and show that both are at least as good as the previous results and significantly better in some scenarios. In particular, we show that our achievable rate can be higher than twice the one from prior work, and that it is close to the upper bound most of the time.

Follow-up research includes closing the gap between achievable and upper bound, and further adapting our coding scheme to time-varying channels according to a small-frequency feedback provided from destination to the relay and from the relay to source, or directly from destination to source.

\appendix 

\begin{IEEEproof}[Proof of Proposition~\ref{prop:recovery}]
    Let us first assume that $\underline{x}(t)$ is erased and prior message packets $\{\underline{m}(t')\}_{t'\in[0:t-1]}$ are known by the relay. We wish to show that, with every non-erased source packet $\underline{x}(t')$, $t' > t$, the relay will be able to decode $\ell'$ symbols of $\underline{m}(t)$. This will imply that, after $k' = T + 1 - N_1 - N_2$ successfully received source packets, the relay is able to fully decode $\underline{m}(t)$. 
    
    In order to see this, consider the sequence of source sub-packets $\{\underline{x}^{(0)}(t)\}_{t\in[0:\infty]}$ produced by interleaving  systematic $[n',k']$-MDS codes as shown in Fig.~\ref{fig:diag_interleaving_source_relay}. Now, note that the set of code symbols $\{m_0(t-k'+1),m_1(t-k'+2), \cdots,m_{k'-1}(t),p_0(t+1),\cdots,p_{r'-1}(t+r')\}$ (for example, the diagonal highlighted in the figure) forms a codeword of the underlying $[n', k']$-MDS code.
    
    Now, recall that, since we assume prior message packets $\{\underline{m}(t')\}_{t'\in[0:t-1]}$ are available, the $k'-1$ message symbols $m_0(t-k'+1),m_1(t-k'+2), \cdots,m_{k'-2}(t-1)$ are already known. In the illustrated example in Fig.~\ref{fig:diag_interleaving_source_relay}, this means that $m_0(0)$, $m_1(1)$ and so on are known, and the only unknown symbol is $m_{k' - 1}(k' - 1)$. 
    
    Further, note that, since one erasure has already occurred at time $t$, then, at most $N_1 - 1$ more erasures may occur in the time slots $[t + 1 : t + N_1]$. However, there are exactly $N_1$ parity symbols, thus, it is guaranteed that we are able to recover at least one parity symbol $p_{i - 1}(t + i)$ from $\underline{x}(t + i)$, for some $i \in [1:N_1]$. Then, it is clear that we are able to recover $\underline{m}_{k' - 1}(t)$ from the first non-erased coded sub-packet after time $t$.
    
    Similarly, message symbol $\underline{m}_{k'- j}$ will be recovered after $j$ source packets are successfully recovered (i.e. not erased). This is because at most $j - 1$ symbols from the underlying $[n', k']$-MDS codes are unknown at time $t$, thus, as soon as $j$ symbols are recovered, $\underline{m}_{k' - j}$ can be recovered.
    
    Now, let us remove the assumption that previous message packets are known: then, instead of perfectly recovering the message symbol $\underline{m}_{k' - j}$, we are able to recover an \textit{estimate} of it, which contains some linear combination of past message symbols, that is, we recover some $\tilde{\underline{m}}_{k' - j} = \underline{m}_{k' - j} + \sum_{i=0}^{k'-j} \alpha_i \underline{m}_{k' - j - i}$. It is easy to see that this meets our definition of estimate, as the interference from past message symbols can be easily removed if we have access to them. 
    
    Finally, note that so far we have discussed one sub-packet, and we have shown that each sub-packet is able to recover \textbf{one} symbol from each non-erased packet. Since our code construction is obtained by simply appending $\ell'$ sub-packets, the relay is able to decode $\ell'$ estimates with every non-erased source packet, until (an estimate of) the message packet is completely decoded.
\end{IEEEproof}

\bibliographystyle{IEEEtran}
\bibliography{eladd.bib}
% %\bibliography{main.bbl}

\end{document}